\documentclass[letterpaper,twocolumn,10pt]{article}
\usepackage{usenix}

\usepackage{xfrac}
\usepackage{array}
\usepackage[caption=false,font=normalsize,labelfont=sf,textfont=sf]{subfig}
\usepackage{amsmath,amsfonts,amsthm,amssymb}
\usepackage{algorithm}
\usepackage{graphicx}
\usepackage{bm}
\usepackage{algpseudocode}
\usepackage{booktabs}
\usepackage{multirow}
\usepackage{colortbl}
\definecolor{Gray}{gray}{0.935}
\newcolumntype{g}{>{\columncolor{Gray}}c}

\algnewcommand\algorithmicforpara{\textbf{for}}
\algnewcommand\algorithmicdoinparallel{\textbf{do in parallel}}
\algdef{S}[FOR]{ForParallel}[1]{\algorithmicforpara\ #1\ \algorithmicdoinparallel}
\algnewcommand\algorithmicforeach{\textbf{for each}}
\algdef{S}[FOR]{ForEach}[1]{\algorithmicforeach\ #1\ \algorithmicdo}
\algnewcommand\algorithmicforeachparallel{\textbf{for each}}
\algdef{S}[FOR]{ForEachParallel}[1]{\algorithmicforeach\ #1\ \algorithmicdoinparallel}
\algdef{SE}{DoParallel}{EndParallel}{\algorithmicdoinparallel}{\textbf{end parallel}}
\algdef{SE}[SUBALG]{Indent}{EndIndent}{}{\algorithmicend\ }\algtext*{Indent}\algtext*{EndIndent}
\renewcommand{\algorithmicrequire}{\textbf{Input:}}
\renewcommand{\algorithmicensure}{\textbf{Output:}}
\newcommand{\Desc}[3]{\Statex\makebox[#1][l]{#2}#3}

\newcommand{\etal}{\textit{et al.}}
\begin{document}
%-------------------------------------------------------------------------------

%don't want date printed
\date{}

\title{\Large \bf Byzantine-Robust Federated Learning Using Generative Adversarial Networks}

%for single author (just remove % characters)
\author{
{\rm Anonymous}\\
Anonymous Institution
% \and
% {\rm Second Name}\\
% Second Institution
% copy the following lines to add more authors
% \and
% {\rm Name}\\
%Name Institution
} % end author

% \author{{\rm Usama Zafar} \and {\rm André Teixeira} \and {\rm Salman Toor}
% \thanks{Usama Zafar, André Teixeira, and Salman Toor are with Department of Information Technology, Uppsala University, 751 05 Uppsala, Sweden. Additionally, Salman Toor is CTO at Scaleout Systems (e-mail: usama.zafar@it.uu.se; andre.teixeira@it.uu.se; salman.toor@it.uu.se).}% <-this % stops a space
% \thanks{The code is available at: \href{https://github.com/SciML-FL/gan-filter}{https://github.com/SciML-FL/gan-filter}}% <-this % stops a space
% }

%for single author (just remove % characters)
\author{
{\rm Usama Zafar}\\
{\rm Uppsala University}\\
{\rm usama.zafar@it.uu.se}\\
\and
{\rm André M. H. Teixeira}\\
{\rm Uppsala University}\\
{\rm andre.teixeira@it.uu.se}\\
\and
{\rm Salman Toor}\thanks{Also CTO at Scaleout Systems.}\\
{\rm Uppsala University}\\
{\rm salman.toor@it.uu.se}\\
} % end author

\maketitle
\begin{abstract}
Federated learning (FL) enables collaborative model training across distributed clients without sharing raw data, but its robustness is threatened by Byzantine behaviors such as data and model poisoning. Existing defenses face fundamental limitations: robust aggregation rules incur error lower bounds that grow with client heterogeneity, while detection-based methods often rely on heuristics (e.g., a fixed number of malicious clients) or require trusted external datasets for validation. We present a defense framework that addresses these challenges by leveraging a conditional generative adversarial network (cGAN) at the server to synthesize representative data for validating client updates. This approach eliminates reliance on external datasets, adapts to diverse attack strategies, and integrates seamlessly into standard FL workflows. Extensive experiments on benchmark datasets demonstrate that our framework accurately distinguishes malicious from benign clients while maintaining overall model accuracy. Beyond Byzantine robustness, we also examine the representativeness of synthesized data, computational costs of cGAN training, and the transparency and scalability of our approach.
\end{abstract}

\section{Introduction}

The success of modern machine learning hinges on access to large and diverse datasets. Applications such as personalized healthcare, finance, and recommendation systems benefit from data-rich models, but centralizing sensitive data raises severe privacy and regulatory concerns. High-profile incidents, including the Facebook–Cambridge Analytica scandal~\cite{cadwalladr_cambridge_2018} and the Equifax breach~\cite{equifax_breach_2017}, illustrate the risks of centralized data storage, while privacy regulations such as the GDPR~\cite{voigt_gdpr_2017} and CCPA~\cite{ccpa_2018} impose strict limits on data collection and sharing. These developments have accelerated the adoption of federated learning (FL)~\cite{mcmahan_communicationefficient_2023}, which enables multiple organizations or devices to collaboratively train models without exposing raw data. Despite its privacy-preserving promise, FL is vulnerable to Byzantine failures, where some clients behave arbitrarily due to faults or adversarial manipulation. Among these, model poisoning~\cite{fang_local_2020, cao_mpaf_2022} has emerged as the most damaging threat, enabling adversaries to degrade overall model accuracy and hinder convergence.

Efforts to defend against Byzantine behavior have followed two broad directions. One approach employs Byzantine-robust aggregation rules such as Krum~\cite{blanchard_machine_2017}, Coordinate-wise Median~\cite{yin_byzantinerobust_2021}, and Trimmed Mean~\cite{yin_byzantinerobust_2021}, which reduce the influence of corrupted updates through statistical filtering. While effective in homogeneous settings, these methods face fundamental limitations: their error guarantees scale poorly with the variance of client updates, restricting robustness in heterogeneous FL environments~\cite{karimireddy_learning_2021, 2023_Allouah_FixingMixingRecipe}. A second direction leverages detection-based strategies that aim to identify and filter malicious clients~\cite{shen_auror_2016, zhang_fldetector_2022, cao_fltrust_2022, zhao_detecting_2022}. These often depend on heuristics such as assumed Byzantine fractions or external validation datasets—assumptions that are impractical in real deployments and misaligned with FL’s privacy objectives. Moreover, static detection rules are brittle against adaptive adversaries whose strategies evolve during training.

Within this second direction, Zhao \etal~\cite{zhao_detecting_2022} introduced a GAN-based defense that validates updates using auxiliary data. While generative approaches show promise, reliance on external validation datasets undermines privacy guarantees and limits deployability in federated settings. To overcome this, we propose a filter-based defense that leverages a conditional generative adversarial network (cGAN) trained solely from the global model itself. Our method requires no client-side or auxiliary datasets, thereby avoiding additional privacy leakage. By generating synthetic inputs tailored to the evolving global model, the defense adapts naturally to shifting attack strategies. Further, it integrates seamlessly into standard FL workflows without modifying client procedures, offering a practical path toward robust, privacy-preserving, and deployable defenses.

\hfill\\
\noindent
\textbf{Research Focus \& Contributions.} 
Our study investigates how decision boundaries evolve in federated learning under adversarial influence, and how synthetic data can be leveraged to probe these boundaries for robust and privacy-preserving defense. Unlike natural training data, our GAN-generated samples are not limited to dense data regions but are distributed across the input space to better distinguish benign from malicious updates, without exposing any client data. While GANs themselves are not new, our contribution lies in leveraging their ability to generate boundary-respecting samples that may appear unrealistic to humans but remain meaningful to the model. This perspective allows us to analyze how poisoned updates shift the global decision surface and to design detection mechanisms that are effective and privacy-preserving. We also evaluate its computational cost and illustrate the spread of synthesized data across decision boundaries using a toy example, which demonstrates how our method supports a robust defense (further elaborated in Section~\ref{sec:defense_formulation}).

To summarize, this paper makes the following key contributions:
\begin{itemize}
    \item \textbf{First GAN-based defense without external data.} We show that GANs can be trained entirely from the global model itself, enabling the server to generate synthetic data for validating client updates without auxiliary datasets.  
    \item \textbf{Decision-boundary informed detection.} By generating samples that respect decision boundaries beyond dense regions, our method provides model-relevant probes that reveal poisoned behavior even if the generated samples are not human-interpretable.  
    \item \textbf{Privacy-preserving and adaptive.} The defense detects malicious contributions without accessing client data and without relying on static thresholds or additional hyperparameters, simplifying deployment and improving robustness against diverse attack strategies.  
    \item \textbf{Robustness and deployability.} Extensive evaluations across datasets and architectures show that our defense reliably distinguishes malicious from benign clients while maintaining model accuracy, with computational costs analyzed to demonstrate practicality for real-world FL systems.  
\end{itemize}

\begin{table}[!t]
    \centering
    \caption{Summary of notations used throughout the paper.\label{table:notations}}
    \begin{tabular}{l p{0.8\linewidth}}
        \toprule[1.5pt]\midrule[1.0pt]
         Symbol & Meaning \\
         \midrule[1.0pt]
         \(\mathcal{C}\) & Set of all clients (\(N = |\mathcal{C}|\)) \\
         \(\mathcal{B}\) & Set of Byzantine clients (\(M = |\mathcal{B}|\)) \\
         \(\mathcal{H}\) & Set of honest clients (\(\mathcal{C} \setminus \mathcal{B}\)) \\
         \(\varepsilon\) & Fraction of adversarial clients (\(\varepsilon = M / N\)) \\
         \(w_t\) & Global model weights at round \(t\) \\
         \(w_{t}^{i}\) & Local model weights of client \(i\) at round \(t\) \\
         \(\mathcal{A}\) & Aggregation rule \\
         \(D_i\) & Local dataset of client \(i\) \\
         \(n \leq N\) & Number of sampled clients per round \\
         \(\eta\) & Learning rate for local training \\
         \(\alpha\) & Dirichlet parameter controlling data heterogeneity \\
         \(\mathcal{G}\) & Generator model of the conditional GAN \\
         \(\mathcal{D}_{w_t}\) & Discriminator of the conditional GAN, parameterized by global model \(w_t\) \\
         \(D_{\text{syn}}\) & Synthetic dataset generated by \(\mathcal{G}\) \\
         \(\beta\) &   
         \begin{minipage}[t]{\linewidth}
         Byzantine aggregation heuristic parameter:
         \begin{itemize}
             \item In \textsc{TrimAvg}, \(\beta\) is the ratio of updates trimmed from each extreme.  
             \item In \textsc{MultiKrum}, \(\beta\) is the assumed fraction of malicious clients, used to determine how many updates to discard.  
         \end{itemize}
         \end{minipage} \\
         \midrule[1.0pt]\bottomrule[1.5pt]
    \end{tabular}
\end{table}

\section{Background and Related Work}
\subsection{Federated Learning Overview}
\label{section:FL}

Federated Learning (FL) \cite{mcmahan_communicationefficient_2023, yang_federated_2019} is a distributed training paradigm that enables collaborative model training across multiple clients, each holding local datasets, without directly sharing data. This paradigm improves privacy by ensuring that raw data remains on-device, while only model updates are communicated.

Formally, a central server \(\mathcal{S}\), coordinates the training process across a set of \(N\) clients, denoted as \(\mathcal{C} = \{1, 2, \ldots, N\}\), where each client \(i\in\mathcal{C}\) possesses a private dataset, \(D_i\). The global objective is to minimize the empirical risk over the union of all datasets:
\[
\min_{w \in \mathbb{R}^d} f(w) = \sum_{i=1}^N \frac{|D_i|}{|D|} f_i(w), \quad f_i(w) = \frac{1}{|D_i|} \sum_{(x,y) \in D_i} \ell(w; x,y),
\]
\noindent where \(w\) denotes model parameters, \(\ell\) is the loss function, and \(|D| = \sum_{i=1}^N |D_i|\). 

Training proceeds iteratively: at each round \(t\), the server samples a subset \(\mathcal{C}_t \subseteq \mathcal{C}\) with \(|\mathcal{C}_t| = n \leq N\) and broadcasts the current model \(w_t\). Each selected client \(i \in \mathcal{C}_t\) performs local updates to produce \(w_t^i\), and the server aggregates them, typically via weighted averaging (FedAvg)~\cite{mcmahan_communicationefficient_2023}:
\[
w_{t+1} = \frac{1}{K}\sum_{i \in \mathcal{C}_t} |D_i|\;w_t^i ,
\]
\noindent where \(K = \sum_{i \in \mathcal{C}_t} |D_i|\). 

% \figurename~\ref{fig:fig1-FL} illustrates this process.
% \begin{figure}[!t]
%     \centering
%     \includegraphics[width=\columnwidth]{figures/fig1-fedml.pdf}
%     \caption{An overview of Federated Learning (FL) framework.}
%     \label{fig:fig1-FL}
% \end{figure}

\subsection{Byzantine Threats in Federated Learning}
\label{section:poisoningattacksinFL}

Federated Learning (FL) is inherently vulnerable to \textit{Byzantine clients}, participants that deviate from honest behavior due to malicious intent or faults. Such adversaries exploit the server’s implicit trust in client updates, where neither training data nor parameter integrity can be directly verified. This vulnerability is exacerbated in heterogeneous settings, where non-IID distributions naturally induce divergence among honest updates, providing camouflage for adversarial perturbations.

Byzantine threats typically manifest as \textit{data poisoning}, where local datasets are tampered with (e.g., label flips or feature manipulations) to propagate misleading gradients~\cite{chen_targeted_2017, sun_data_2020}, or as \textit{model poisoning}, where adversaries directly craft malicious parameter updates to bias or destabilize the global model~\cite{2019_Bagdasaryan_HowBackdoorFederated, 2020_Fang_LocalModelPoisoning}. Attacks may be \textit{untargeted}, aiming to degrade overall accuracy, or \textit{targeted}, enforcing specific misclassifications. Their subtlety, often designed to mimic the variance of benign updates, poses a central challenge for detection, particularly against adaptive adversaries that evolve to evade robust aggregation mechanisms.

%%%%%%%%%%%%%%%%%%%%%%%%%%%%%%%%%%%%%%%%%%%%%
\begin{figure}[!t]
    \centering
    \includegraphics[width=\linewidth]{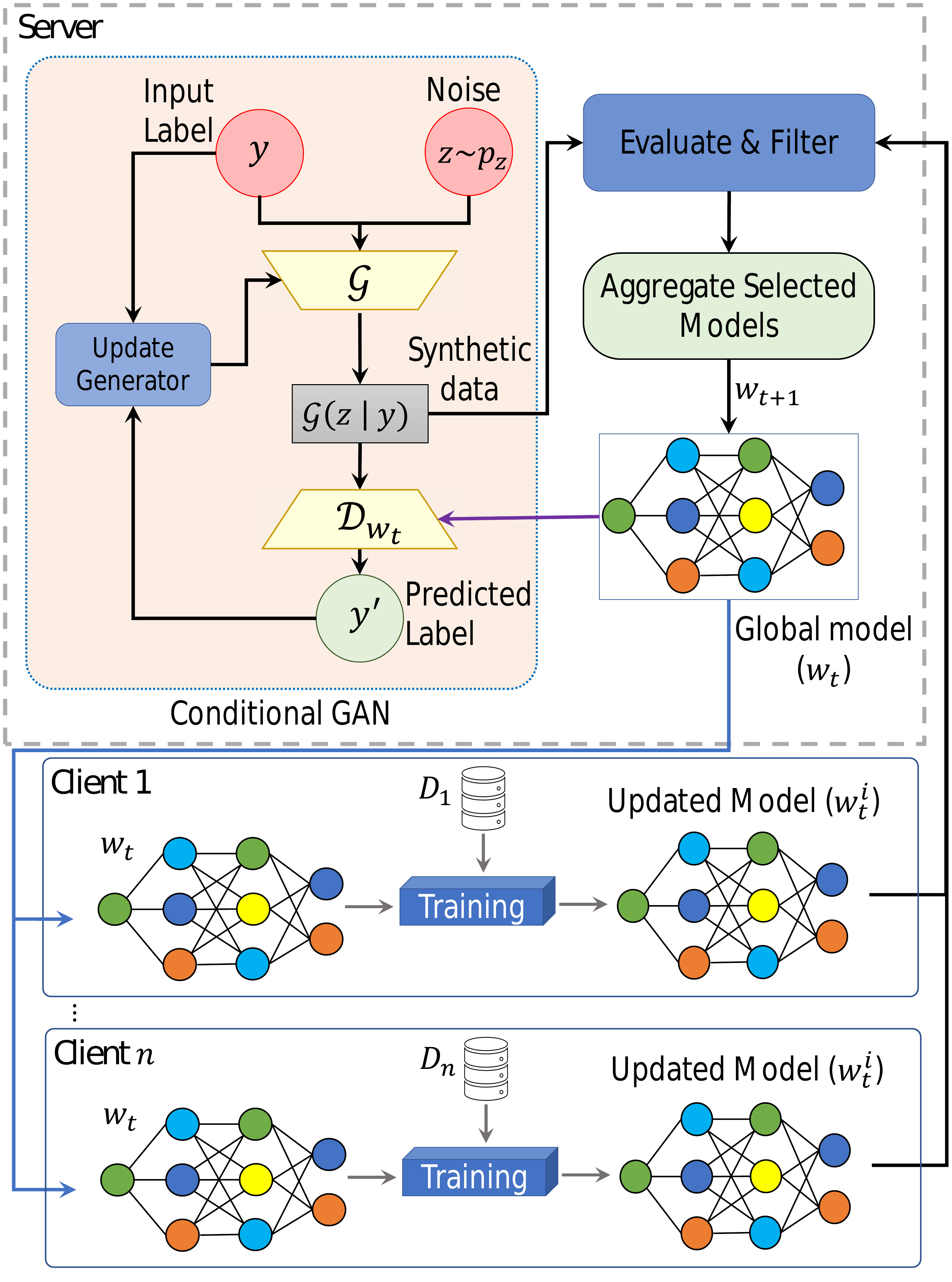}
    \caption{Proposed defense pipeline to authenticate updates.}
    \label{fig:fig2-framework}
\end{figure}
%%%%%%%%%%%%%%%%%%%%%%%%%%%%%%%%%%%%%%%%%%%%%

\subsection{Defenses Against Byzantine Attacks}
\label{section:defenseagainstpoisoningattacks}

The problem of Byzantine threats in FL has received significant attention in the recent years and a number of defenses have been proposed with various advantages and drawbacks. These defenses fall into two broad categories: \textit{Robust aggregation mechanisms}, and \textit{Filtration-based techniques}.

\textit{\textbf{Robust aggregation mechanisms}} aim to reduce the impact of malicious updates during model aggregation by designing the Byzantine-robust aggregation rules that are resilient to outliers and adversarial manipulations. Popular examples include \texttt{Coordinate-wise Median} and \texttt{Trimmed-Mean}~\cite{yin_byzantinerobust_2021}~\cite{yin_byzantinerobust_2021, li_experimental_2023, chen_distributed_2017}, which respectively down-weight or exclude extreme coordinate values; \texttt{Krum}~\cite{blanchard_machine_2017}, which selects the update closest to the majority; \texttt{Bulyan}\cite{mhamdi_hidden_2018}, which combines Krum and Trimmed-Mean; and \texttt{RFA}\cite{pillutla_robust_2022}, which aggregates via the geometric median. While these rules improve resilience to outliers, their effectiveness degrades under large-scale collusion or adaptive attacks. Moreover, they are subject to fundamental error lower bounds~\cite{karimireddy_learning_2021, 2023_Allouah_FixingMixingRecipe}, which we discuss in Section~\ref{section:limitationsofexistingdefenses}.

\textit{\textbf{Filtration-based techniques}} instead attempt to detect and exclude malicious updates before aggregation. One line of work uses clustering (e.g., \texttt{Auror}\cite{shen_auror_2016}, \texttt{FoolsGold}\cite{fung_limitations_2020}, \texttt{Flame}\cite{nguyen_flame_2023}) to identify anomalous updates, assuming benign clients form a coherent cluster. Another line relies on validation-based authentication, where updates are scored against a reference dataset (e.g., \texttt{FLTrust}\cite{cao_fltrust_2022}, \texttt{GAN-based validation}~\cite{zhao_detecting_2022}). More recent approaches (e.g., \texttt{DeepSight}\cite{rieger_deepsight_2022}, \texttt{FedDMC}\cite{mu_feddmc_2024}) extend these ideas with adversarial feature analysis or deep anomaly detection. While effective, filtration methods often depend on clean validation data or incur higher computational costs.

%%%%%%%%%%%%%%%%%%%%%%%%%%%%%%%%%%%%%%%%%%%%%%%%%%%%%%%%%%%%
\begin{figure*}[!t]
    \centering
    \includegraphics[width=0.98\linewidth]{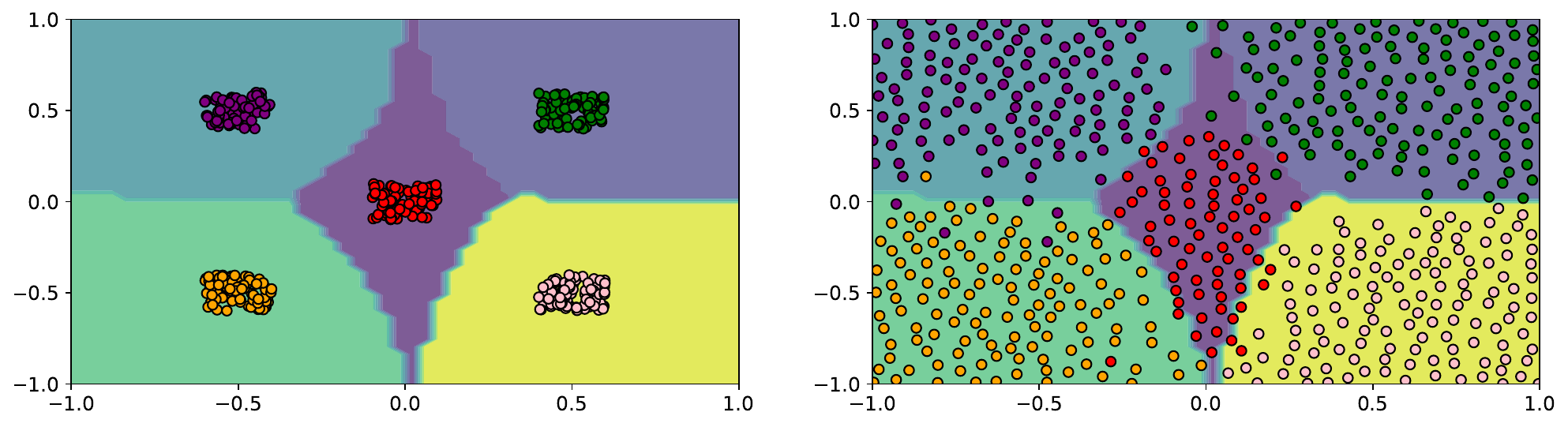}
    \caption{A 2D toy illustration of our approach. \textbf{Left:} the original dataset (points) and decision boundaries learned by a simple MLP. \textbf{Right:} synthetic data generated by our method, filtered so that each point is at least a minimum distance apart for clarity.}
    \label{fig:fig3-decision_boundaries}
\end{figure*}
%%%%%%%%%%%%%%%%%%%%%%%%%%%%%%%%%%%%%%%%%%%%%%%%%%%%%%%%%%%%

\subsection{Limitations of Byzantine Defenses}
\label{section:limitationsofexistingdefenses}

Despite significant progress, existing defenses against Byzantine attacks in federated learning face fundamental limitations that constrain their robustness and scalability. Broadly, these limitations arise from both \emph{practical trade-offs} in algorithm design and \emph{theoretical lower bounds} that no aggregation-based defense can overcome.

From a practical perspective, aggregation rules often incur accuracy or efficiency penalties. For instance, \texttt{Krum} selects only one update per round, which significantly slows convergence~\cite{blanchard_machine_2017}. Its variant, \texttt{MultiKrum}, requires prior knowledge of the number of malicious clients, an unrealistic assumption in practice. \texttt{Trimmed Mean} depends on heuristics to determine how many extreme values to discard, which can lead to under- or over-pruning depending on the distribution of updates~\cite{yin_byzantinerobust_2021}. Filtration-based defenses such as \texttt{FLTrust}~\cite{cao_fltrust_2022} or the work by Li~\etal~\cite{li_learning_2020} improve robustness by leveraging a clean reference dataset, but this reliance on external data limits scalability and privacy. Generative-data-based defenses, such as Zhao~\etal~\cite{zhao_detecting_2022}, similarly use GANs for validation but still depend on centralized or external datasets for training.

Beyond these implementation-level trade-offs, there are also \emph{irreducible theoretical limits} on the robustness of aggregation-based defenses. Recent results show that for any permutation-invariant aggregation rule, the expected optimization error under an \(\varepsilon\) fraction of Byzantine clients is bounded below by
\[
\mathbb{E}[f(\hat{w}_t)] - f(w^\star) \;\; \geq \; \Omega\!\left(\frac{\varepsilon \sigma^2}{\mu}\right),
\]
where \(\mu\) denotes the strong convexity parameter and \(\sigma^2\) the variance of the gradients~\cite{2021_Karimireddy_LearningHistoryByzantine, 2023_Karimireddy_ByzantineRobustLearningHeterogeneous}. In heterogeneous data settings, the error can also be expressed as
\[
\mathbb{E}\!\left[\|\hat{w} - \mu\|^2\right] \;\; \geq \; \varepsilon \rho^2,
\]
with \(\rho^2\) bounding pairwise distances between honest updates. Strikingly, if the adversarial fraction \(\varepsilon \geq 1/2\), the error may even become unbounded, fundamentally limiting the efficacy of robust aggregation in highly adversarial regimes. These results demonstrate that robust-aggregation alone cannot circumvent a fundamental accuracy barrier.

Taken together, these findings underscore the inherent limitations of aggregation-centric defenses. Filtration-based approaches attempt to overcome these constraints by detecting and removing malicious updates, but they often depend on clean reference datasets or trusted validation, which can limit scalability and practicality. This combination of aggregation lower bounds and filtration dependencies motivates the exploration of alternative strategies.

In this work, we propose a generative-data-driven filtration mechanism that eliminates the need for external datasets while effectively mitigating the fundamental accuracy limits imposed on aggregation-based defenses. Specifically, by training a generator on the global model itself, our method leverages synthetic data to validate updates, circumventing the error lower bounds associated with aggregation-only rules while preserving privacy.

\section{Problem Setup}
In this section, we define the threat model considered in our work, describe the defense objectives guiding our framework, and specify the assumptions made.

\subsection{Threat Model}
We consider the classical FL setup introduced in Section~\ref{section:FL}, where \(N\) clients collaboratively train a global model with parameters \(w \in \mathbb{R}^d\) under the coordination of a central server \(\mathcal{S}\). A fraction \(\varepsilon\) of clients, denoted \(\mathcal{B} \subseteq \mathcal{C}\), are controlled by an adversary, so that \(|\mathcal{B}| = \varepsilon N = M\).

The adversary aims to disrupt the training process by injecting malicious updates into the aggregation process. We assume a \textbf{non-omniscient} adversary with the following capabilities:
\begin{itemize}
    \item \textbf{Local knowledge:} The adversary has access to the data, local updates, and training hyperparameters of compromised clients.
    \item \textbf{Aggregation awareness:} The adversary knows the server’s aggregation rule and any deployed defense mechanism.
    \item \textbf{Limited visibility:} The adversary cannot access the raw data of honest clients, though it may observe their updates in certain adaptive attack scenarios.
    \item \textbf{Collusion:} Compromised clients may coordinate their strategies and leverage their own local datasets, similar to honest clients.
\end{itemize}
We assume the server behaves honestly, and compromised clients can manipulate only their own data and updates. This threat model captures realistic adversarial scenarios in federated learning, where malicious clients may collude yet cannot compromise honest participants or the server.

\subsection{Defense Objectives}
Motivated by the limitations of existing aggregation- and filtration-based defenses (Section~\ref{section:limitationsofexistingdefenses}), our framework is designed to achieve the following objectives:
\begin{enumerate}
\item \textbf{Robustness:} Preserve global model integrity under poisoning attacks, including adaptive and intermittent strategies.
\item \textbf{Generality:} Operate without prior knowledge of the number of adversaries \(M\) or the specific attack strategies.
\item \textbf{Fairness:} Minimize false rejection of benign client updates while preserving decision-boundary integrity
\end{enumerate}
These objectives are achieved through a dynamic, synthetic-data-driven mechanism that probes the global model’s decision boundaries to detect malicious updates. The following section formalizes our defense framework, showing how boundary-aligned synthetic samples, generated without accessing client data, enable robustness, generality, and fairness in federated learning.

\section{Defense Formulation}
\label{sec:defense_formulation}

Our defense builds on the intuition that machine learning classifiers learn decision boundaries to separate data points into classes, while adversaries in poisoning attacks attempt to perturb these boundaries to degrade model performance. Effective defenses must therefore detect or neutralize such perturbations. Authentication-based methods provide a promising direction by validating client updates against a trusted reference; however, existing approaches require clean validation datasets, which is unrealistic in privacy-preserving FL.

We address this problem by re-purposing conditional Generative Adversarial Networks (cGANs)~\cite{mirza2014conditional} to generate synthetic data aligned with the decision geometry of the global model, rather than replicating private distributions. Unlike prior GAN-based defenses that rely on auxiliary datasets, our method produces boundary-focused samples that may appear unrealistic to humans but remain meaningful to the model for detecting poisoned updates. Figure~\ref{fig:fig3-decision_boundaries} illustrates this principle using a toy example: real data is clustered, whereas synthetic samples spread across the decision space along critical decision boundaries, providing informative probes for identifying malicious contributions.

To achieve this, we introduce three key modifications: (i) replacing the discriminator with the global model, (ii) training only the generator while keeping the global model fixed, and (iii) eliminating real/fake samples entirely, enabling fully data-free training. These changes yield boundary-aligned synthetic data that is independent of client distributions and serves as a trusted reference for update authentication. This approach directly addresses the robustness, generality, and fairness objectives by providing meaningful probes for detecting malicious contributions without exposing client data. 

While more recent conditional GAN variants (e.g., ccGAN~\cite{2020_Ding_CcGANContinuousConditional}) could be explored, we use a standard cGAN in this work. This choice reflects both the preliminary nature of our study, focusing on validating the core defense concept, and practical considerations: standard cGANs provide a favorable balance between training complexity and computational resources, making them well-suited for large-scale federated experiments. In the next subsection, we detail the training procedure for the generator, showing how these modifications produce boundary-aligned synthetic data suitable for update authentication.

%%%%%%%%%%%%%%%%%%%%%%%%%%%%%%%%%%%%%%%%%%%%%%%%%%%%%%%%%%%%
\begin{algorithm}[!t]
	\caption{Federated Learning with GAN-based Defense}\label{alg:FL_CGAN}
	\begin{algorithmic}[1]
        \renewcommand{\algorithmicrequire}{\textbf{Input:}}
        \renewcommand{\algorithmicensure}{\textbf{Output:}}
        \Require 
            \Desc{3em}{\(\eta\)}{Learning rate}
            \Desc{3em}{\(T\)}{Total communication rounds}
            \Desc{3em}{\(E\)}{Number of local epochs}
            \Desc{3em}{\(B\)}{Batch size}
            \Desc{3em}{\(\mathcal{A}\)}{Server-side aggregation rule}
		\Ensure
            \Desc{3em}{\(w_T\)}{Trained global model}
        \State \(w_0 \gets\) random initialization
        \For{\(t = 1, 2, \dots, T\)}
            \State Randomly samples a set of \(n\) clients \(\mathcal{C}_t\)
            \State Send \(w_{t-1}\) to all sampled clients
            \State \textbf{Client-side (in parallel):}
            \Indent
                \ForEach{\(i \in \mathcal{C}_t\)}
                    \State \(w_t^{i} \gets \text{LocalTraining}(i, w_{t-1}, \eta, B, E)\)
                    \State Send \(w_t^{i}\) to server
                \EndFor
            \EndIndent
            \State \textbf{Server-side (in parallel):}
            \Indent
                \State Initialize discriminator \(\mathcal{D}_{w_{t-1}}\)
                \State Train generator \(\mathcal{G}\) of cGAN
                \State Generate synthetic dataset \(D_{\text{syn}}\) using \(\mathcal{G}\)
            \EndIndent
            \State Collect updates \(\{w_t^{i}\}_{i\in\mathcal{C}_t}\)
            \State \(S_t \gets \text{FilterUpdates}(\{w_t^{i}\}, D_{\text{syn}})\)
            \State \(w_{t} \gets \mathcal{A}(\{w_{t}^{j}\}_{j \in S_t})\)
        \EndFor\\
        \Return \(w_{T}\) \Comment{Final global model}
    \end{algorithmic}
\end{algorithm}
%%%%%%%%%%%%%%%%%%%%%%%%%%%%%%%%%%%%%%%%%%%%%%%%%%%%%%%%%%%%

\subsection{Training the Generator}
Unlike conventional cGANs, our generator \(\mathcal{G}\) is trained not to deceive a binary discriminator, but to align with the decision boundaries of the global model \(w_t\), ensuring synthetic data reflects the model’s current knowledge. Formally, given noise \(z \sim p_z\) and labels \(y \sim p_y\), \(\mathcal{G}\) produces synthetic samples \(\mathcal{G}(z|y)\), which are evaluated by the global model \(\mathcal{D}_{w_t}\). The generator is updated via the cross-entropy (CE) loss:
\begin{equation}
\min_{\mathcal{G}}\mathcal{L}_{\mathcal{G}} = \mathbb{E}_{{z\sim p_z, y\sim p_y}} \big[ \ell(\mathcal{D}_{w_{t}}(\mathcal{G}(z|y)), \;y)\big] \label{eq:gan_objective}
\end{equation}
Training proceeds by sampling \((z,y)\), generating \(\mathcal{G}(z|y)\), evaluating \(\mathcal{D}_{w_t}(\mathcal{G}(z|y))\), and updating \(\mathcal{G}\) by minimizing CE loss. This ensures that generated samples remain aligned with the classifier’s evolving decision regions.

%%%%%%%%%%%%%%%%%%%%%%%%%%%%%%%%%%%%%%%%%%%%%%%%%%%%%%%%%%%%
\begin{algorithm}[!t]
	\caption{Local model training of client \(i\)}\label{alg:trainmodel}
	\begin{algorithmic}[1]
        \Require
            \Desc{3em}{\(w_t\)}{Global model}
            \Desc{3em}{\(\eta\)}{Learning rate}
            \Desc{3em}{\(B\)}{Batch size}
            \Desc{3em}{\(E\)}{Number of local epochs}
        \Ensure 
            \Desc{3em}{\(w_t^{i}\)}{Updated local model}
        \State \(w_t^{i} \gets w_t\)
        \If{\(i \in \mathcal{B}\)} \Comment{Client \(i\) is malicious}
            \State \(w_t^{i} \gets \text{PoisonAttack}(w_t^{i})\)
        \Else \Comment{Client \(i\) is benign}
            \For{\(e = 1, \dots, E\)}
                \State Sample mini-batch \(D \subset D_i\) of size \(B\)
                \State \(w_t^{i} \gets w_t^{i} - \eta \nabla \ell(D, w_t^{i})\)
            \EndFor
        \EndIf\\
        \Return \(w_t^{i}\)
    \end{algorithmic}
\end{algorithm}

\begin{algorithm}[!t]
	\caption{Filtering client updates}\label{alg:filter_updates}
	\begin{algorithmic}[1]
        \Require
            \Desc{5em}{\(m\)}{Filter (\texttt{Fixed}, \texttt{Adaptive}, \texttt{Cluster})}
            \Desc{5em}{\(\{w_t^i\}\)}{Set of client models}
            \Desc{5em}{\(D_{\text{syn}}\)}{Synthetic validation dataset}
            \Desc{5em}{\(\text{EVAL}(\cdot)\)}{Evaluation function}
            \Desc{5em}{\(\tau\)}{(optional) Threshold for \texttt{Fixed} method}
        \Ensure
            \Desc{3em}{\(S_t\)}{Set of accepted client models}
        \State Compute metrics: 
            \(\text{metrics}[w_t^i] \gets \text{EVAL}(w_t^i, D_{\text{syn}}), \;\forall i\)
                \State Transform to scores (so that \emph{higher = better}):
            \[\text{score}[w] = 
              \begin{cases}
                -\text{metrics}[w], & \text{if metric is loss} \\
                +\text{metrics}[w], & \text{if metric is accuracy}
              \end{cases}\]
        \State Initialize \(S_t \gets \emptyset\)
        \If{\(m = \texttt{Fixed}\)}
            \State \(S_t \gets \{\, w_t^i \mid \text{score}[w_t^i] > \tau \,\}\)
        \ElsIf{\(m = \texttt{Adaptive}\)}
            \State \(\tau \gets \text{Mean}(\text{score})\)
            \State \(S_t \gets \{\, w_t^i \mid \text{score}[w_t^i] > \tau \,\}\)
        \ElsIf{\(m = \texttt{Cluster}\)}
            \State Partition scores: \(C_1, C_2 \leftarrow \text{K-Means}(\text{score}, 2)\)
            \State Identify best client: \(w^+ \gets \arg\max_i \text{score}[w_t^i]\)
            \State \(S_t \gets C_j \;\;\text{s.t.}\;\; w^+ \in C_j\) \Comment{Cluster with best client}
        \EndIf\\
        \Return \(S_t\)
    \end{algorithmic}
\end{algorithm}

%%%%%%%%%%%%%%%%%%%%%%%%%%%%%%%%%%%%%%%%%%%%%%%%%%%%%%%%%%%%

\subsection{Generating Synthetic Data}
\label{sec:gen_data}
After training, the generator produces a synthetic validation set
\[
D_{\text{syn}} = \{(X_i, Y_i)\}_{i=1}^K,
\]
with $K = q \cdot C$, where $C$ is the number of classes and $q$ is the number of per-class samples. Although these synthetic points may not resemble raw inputs, they populate the decision boundaries that are most susceptible to adversarial manipulation, making them effective for validating model updates.

\subsection{Authentication of Model Updates}
The server uses $D_{\text{syn}}$ as a trusted reference to authenticate client updates. Each update $w_t^i$ is evaluated on $D_{\text{syn}}$ using a metric $\mathcal{F}$ (e.g., accuracy or loss). Updates are then filtered using either threshold-based or clustering-based methods.  

In the fixed threshold variant, $w_t^{i}$ is accepted if $\mathcal{F}(w_t^{i}; D_{\text{syn}}) > \tau$, where $\tau$ is a constant. Adaptive thresholds adjust $\tau_t$ dynamically (e.g., setting it to the mean or median of the current round’s metrics), improving resilience against evolving attacks. Alternatively, clustering-based methods (e.g., K-Means) group updates by their performance on $D_{\text{syn}}$ and flag outliers as malicious.  

Algorithm~\ref{alg:filter_updates} summarizes the filtering process. While our evaluation shows adaptive thresholds provide the best trade-off, the framework is agnostic: any authentication method requiring a validation set can operate on $D_{\text{syn}}$, without assuming access to clean external data.

\section{Experimental Setting}
We evaluate our framework on widely used image classification datasets, implement several representative poisoning attacks, compare against state-of-the-art baselines, and report standard evaluation metrics. All attacks and defenses are implemented according to their original designs, using provided code when available.

\subsection{Datasets and Model}
We consider three benchmark datasets:
\begin{itemize}
    \item \textbf{MNIST}\cite{deng2012mnist}: 28×28 grayscale handwritten digits (60,000 training, 10,000 test samples).
    \item \textbf{Fashion-MNIST (FMNIST)}\cite{xiao2017fmnist}: 28×28 grayscale images of clothing items (60,000 training, 10,000 test samples).
    \item \textbf{CIFAR-10}~\cite{krizhevsky2009learning}: 32×32 RGB images across 10 classes (50,000 training, 10,000 test samples).
\end{itemize}
\noindent 
We use LeNet-5~\cite{lecun1998gradient} for MNIST and FMNIST, and ResNet-18~\cite{he2015deep} for CIFAR-10. To simulate statistical heterogeneity, data is partitioned among clients using a class-wise Dirichlet distribution~\cite{zhang_flip_2023,2021_Karimireddy_LearningHistoryByzantine}, with concentration parameter \(\alpha\). We evaluate four regimes: (i) near-IID (\(\alpha=100\)), (ii) low non-IID (\(\alpha=10\)), (iii) moderate non-IID (\(\alpha=1\)), and (iv) extreme non-IID (\(\alpha=0.1\)). Results for the near-IID case (\(\alpha=100.0\)) and the extreme non-IID case (\(\alpha=0.1\)) are presented in the main body, while intermediate settings (\(\alpha=10.0, 1.0\)) are reported in Appendix~\ref{sec:appendix-b}.

\subsection{System Setup}
We consider \(100\) clients in total, of which a subset may be malicious. In each communication round, \(10\) clients are randomly selected to perform \(10\) local epochs of training using SGD with a learning rate of \(0.01\) and a batch size of \(128\). The updated models are then submitted to the server for aggregation. This process is repeated for \(200\) global communication rounds. Additional hyperparameters and setup details are provided in Appendix~\ref{section:appendix-a}.

\subsection{Attack Setup}
\label{sec:attack_setup}
To evaluate the robustness of our framework, we implement the following poisoning attacks:
\begin{itemize}
    \item \textbf{Random Noise (RN)}: Malicious clients submit random updates sampled from a normal distribution.
    \item \textbf{Sign Flipping (SF)}~\cite{karimireddy_learning_2021}: Clients invert the sign of their gradient updates, effectively performing gradient ascent.
    \item \textbf{Label-Flipping (LF)}~\cite{fang_local_2020}: Clients flip the labels of their local training data to degrade global model performance.
    \item \textbf{Inner Product Manipulation (IPM)}: Clients manipulate gradient directions to maximize divergence from the benign average.
\end{itemize}
We vary the malicious client proportion in \(\{10\%, 20\%, 30\%\}\), consistent with prior federated learning threat models~\cite{shejwalkar_back_2021}. By default, adversaries behave maliciously in every selected round.

\subsection{Compared Baselines}
We compare our method against widely recognized aggregation strategies:
\begin{itemize}
    \item \textbf{\textsc{FedAvg}}~\cite{mcmahan_communicationefficient_2023}: Standard averaging of client updates.
    \item \textbf{\textsc{Median}}~\cite{yin_byzantinerobust_2021}: Coordinate-wise median to mitigate outliers.
    \item \textbf{Geometric Median (\textsc{GeoMedian})}~\cite{pillutla_robust_2022}: Aggregation via geometric median for Byzantine robustness.
    \item \textbf{Trimmed Mean (\textsc{TrimAvg})}~\cite{yin_byzantinerobust_2021}: Removes extreme updates before averaging.
    \item \textbf{\textsc{MultiKrum}}~\cite{blanchard_machine_2017}: Selects updates most similar to others for aggregation.
    \item \textbf{Nearest Neighbor Mixing (\textsc{NNM+Krum})}~\cite{2023_Allouah_FixingMixingRecipe}: Combines nearest-neighbor filtering with Krum aggregation.
    \item \textbf{Nearest Neighbor Mixing (\textsc{NNM+Krum})}~\cite{2023_Allouah_FixingMixingRecipe}: A two-step robust aggregation method that first smooths client updates using nearest-neighbor averaging to mitigate the effect of outliers, and then applies Krum to select the update closest to the majority.
    % \item \textbf{FLTrust}~\cite{cao_fltrust_2022}: A trust-based aggregation method that uses a root dataset to evaluate the reliability of client updates.
\end{itemize}
These baselines cover both statistical heterogeneity and adversarial robustness, providing a comprehensive set of comparisons.

\subsection{Evaluation Metrics}
\label{sec:eval_metrics}
We report the following metrics:
\begin{itemize}
    \item \textbf{Main Task Accuracy (ACC)}: Accuracy of the global model on a clean test set, reflecting primary task performance.
    \item \textbf{True Positive Rate (TPR)}: Fraction of malicious updates correctly identified:
    \[
        \text{TPR} = \frac{\text{TP}}{\text{TP} + \text{FN}},
    \]
    where TP is the number of malicious models correctly detected, and FN the number missed.
    \item \textbf{True Negative Rate (TNR)}: Fraction of benign updates correctly identified:
    \[
        \text{TNR} = \frac{\text{TN}}{\text{TN} + \text{FP}},
    \]
    where TN is the number of benign models correctly accepted, and FP the number falsely flagged as malicious.
\end{itemize}
These metrics together measure task accuracy, attack resilience, and detection effectiveness~\cite{nguyen_flame_2022,2023_Allouah_FixingMixingRecipe}.

%%%%%%%%%%%%%%%%%%%%%%%%%%%%%%%%%%%%%%%%%%%%%%%%%%%%%%%%%%%%%%%%%%%%%%%%%%%%%%%%
\begin{table*}[!ht]
    % {\centering
    \begin{center}
        \caption{\label{tab:results-alpha-100} Overall test accuracy (ACC) of federated learning under benign training and poisoning attacks across malicious client fractions (\(10\%\), \(20\%\), \(30\%\)) in the near-IID setting \(\alpha = 100\). Results are reported on MNIST, Fashion-MNIST, and CIFAR-10, comparing standard aggregation rules (\textsc{FedAvg, Median, Trimmed Mean, GeoMedian, Krum, NNM+Krum}) against our proposed GAN-based authentication variants. GAN-based defenses consistently preserve accuracy even under high adversarial presence, whereas conventional robust aggregation methods degrade significantly.}
        
        \resizebox{1.0\linewidth}{!}{%
            \begin{tabular}{ll g cgcg cgcg cgcg}\toprule[1.5pt]\midrule[1.125pt]
                & \multirow{2}{*}{\textbf{Baseline}} & \multicolumn{1}{c}{\multirow{2}{*}{\parbox{1cm}{\centering No Attack}}} & \multicolumn{4}{c}{$\epsilon = 10\%$} & \multicolumn{4}{c}{$\epsilon = 20\%$} & \multicolumn{4}{c}{$\epsilon = 30\%$} \\
                \cmidrule(lr){4-7} \cmidrule(lr){8-11} \cmidrule(lr){12-15}
                &  & \multicolumn{1}{c}{} & \multicolumn{1}{c}{\textbf{RN}} & \multicolumn{1}{c}{\textbf{LF}}  & \multicolumn{1}{c}{\textbf{SF}}  & \multicolumn{1}{c}{\textbf{IPM}} & \multicolumn{1}{c}{\textbf{RN}} & \multicolumn{1}{c}{\textbf{LF}}  & \multicolumn{1}{c}{\textbf{SF}}  & \multicolumn{1}{c}{\textbf{IPM}} & \multicolumn{1}{c}{\textbf{RN}} & \multicolumn{1}{c}{\textbf{LF}}  & \multicolumn{1}{c}{\textbf{SF}}  & \multicolumn{1}{c}{\textbf{IPM}} \\
				\midrule[1.125pt]
				\multirow{8}{*}{\rotatebox{90}{\textbf{MNIST}}}
				  & \textsc{FedAvg} & 0.99 & 0.88 & 0.98 & 0.11 & 0.11 & 0.54 & 0.11 & 0.11 & 0.11 & 0.10 & 0.11 & 0.10 & 0.10 \\
				~ & \textsc{Median} & 0.99 & 0.99 & 0.99 & 0.99 & 0.99 & 0.90 & 0.11 & 0.99 & 0.10 & 0.10 & 0.11 & 0.11 & 0.10 \\
				~ & \textsc{TrimAvg} & 0.99 & 0.91 & 0.99 & 0.98 & 0.11 & 0.85 & 0.97 & 0.95 & 0.10 & 0.10 & 0.11 & 0.11 & 0.10 \\
				~ & \textsc{GeoMedian} & 0.99 & 0.99 & 0.99 & 0.99 & 0.99 & 0.96 & 0.98 & 0.99 & 0.10 & 0.10 & 0.11 & 0.11 & 0.10 \\
				~ & \textsc{MultiKrum} & 0.99 & 0.94 & 0.99 & 0.98 & 0.11 & 0.40 & 0.98 & 0.11 & 0.10 & 0.10 & 0.11 & 0.11 & 0.10 \\
				~ & \textsc{NNM+Krum} & 0.99 & 0.93 & 0.99 & 0.98 & 0.11 & 0.89 & 0.95 & 0.11 & 0.10 & 0.10 & 0.11 & 0.11 & 0.10 \\ \cmidrule(lr){2-15}
				~ & \textsc{GAN (Fixed Threshold)} & 0.99 & 0.99 & 0.99 & 0.99 & 0.99 & 0.99 & 0.99 & 0.99 & 0.99 & 0.99 & 0.99 & 0.99 & 0.99 \\
				~ & \textsc{GAN (Adaptive: Mean)} & 0.99 & 0.99 & 0.99 & 0.99 & 0.99 & 0.99 & 0.99 & 0.99 & 0.99 & 0.99 & 0.99 & 0.99 & 0.99 \\
				~ & \textsc{GAN (Clustered)} & 0.99 & 0.99 & 0.99 & 0.99 & 0.99 & 0.99 & 0.99 & 0.99 & 0.99 & 0.99 & 0.99 & 0.99 & 0.99 \\
				\midrule[1.125pt]
				\multirow{8}{*}{\rotatebox{90}{\textbf{Fashion MNIST}}}
				  & \textsc{FedAvg} & 0.89 & 0.71 & 0.89 & 0.10 & 0.10 & 0.27 & 0.86 & 0.10 & 0.10 & 0.10 & 0.10 & 0.10 & 0.10 \\
				~ & \textsc{Median} & 0.89 & 0.89 & 0.88 & 0.88 & 0.89 & 0.76 & 0.80 & 0.87 & 0.10 & 0.10 & 0.10 & 0.10 & 0.10 \\
				~ & \textsc{TrimAvg} & 0.89 & 0.66 & 0.88 & 0.79 & 0.10 & 0.69 & 0.81 & 0.10 & 0.10 & 0.10 & 0.10 & 0.10 & 0.10 \\
				~ & \textsc{GeoMedian} & 0.89 & 0.88 & 0.89 & 0.88 & 0.89 & 0.84 & 0.86 & 0.87 & 0.10 & 0.10 & 0.10 & 0.10 & 0.10 \\
				~ & \textsc{MultiKrum} & 0.89 & 0.78 & 0.89 & 0.10 & 0.10 & 0.71 & 0.88 & 0.10 & 0.10 & 0.10 & 0.10 & 0.10 & 0.10 \\
				~ & \textsc{NNM+Krum} & 0.89 & 0.77 & 0.89 & 0.10 & 0.10 & 0.72 & 0.87 & 0.10 & 0.10 & 0.10 & 0.10 & 0.10 & 0.10 \\ \cmidrule(lr){2-15}
				~ & \textsc{GAN (Fixed Threshold)} & 0.89 & 0.89 & 0.89 & 0.89 & 0.89 & 0.89 & 0.89 & 0.89 & 0.89 & 0.88 & 0.88 & 0.88 & 0.88 \\
				~ & \textsc{GAN (Adaptive: Mean)} & 0.89 & 0.89 & 0.89 & 0.89 & 0.89 & 0.89 & 0.89 & 0.89 & 0.89 & 0.89 & 0.89 & 0.88 & 0.89 \\
				~ & \textsc{GAN (Clustered)} & 0.89 & 0.89 & 0.89 & 0.89 & 0.89 & 0.89 & 0.89 & 0.89 & 0.89 & 0.89 & 0.89 & 0.88 & 0.89 \\
				\midrule[1.125pt]
				\multirow{8}{*}{\rotatebox{90}{\textbf{CIFAR-10}}}
				  & \textsc{FedAvg} & 0.88 & 0.12 & 0.88 & 0.59 & 0.10 & 0.11 & 0.86 & 0.10 & 0.10 & 0.11 & 0.60 & 0.10 & 0.10 \\
				~ & \textsc{Median} & 0.88 & 0.87 & 0.88 & 0.86 & 0.86 & 0.44 & 0.86 & 0.82 & 0.10 & 0.13 & 0.83 & 0.72 & 0.10 \\
				~ & \textsc{TrimAvg} & 0.89 & 0.13 & 0.87 & 0.82 & 0.27 & 0.12 & 0.86 & 0.75 & 0.44 & 0.13 & 0.84 & 0.63 & 0.10 \\
				~ & \textsc{GeoMedian} & 0.89 & 0.88 & 0.88 & 0.87 & 0.87 & 0.43 & 0.87 & 0.84 & 0.10 & 0.13 & 0.86 & 0.76 & 0.10 \\
				~ & \textsc{MultiKrum} & 0.89 & 0.13 & 0.88 & 0.82 & 0.10 & 0.15 & 0.88 & 0.79 & 0.17 & 0.15 & 0.85 & 0.76 & 0.30 \\
				~ & \textsc{NNM+Krum} & 0.89 & 0.13 & 0.88 & 0.82 & 0.10 & 0.13 & 0.87 & 0.80 & 0.18 & 0.12 & 0.84 & 0.77 & 0.10 \\ \cmidrule(lr){2-15}
				~ & \textsc{GAN (Fixed Threshold)} & 0.56 & 0.58 & 0.56 & 0.56 & 0.56 & 0.57 & 0.57 & 0.57 & 0.57 & 0.56 & 0.57 & 0.56 & 0.57 \\
				~ & \textsc{GAN (Adaptive: Mean)} & 0.88 & 0.88 & 0.88 & 0.88 & 0.88 & 0.88 & 0.88 & 0.88 & 0.88 & 0.87 & 0.87 & 0.88 & 0.88 \\
				~ & \textsc{GAN (Clustered)} & 0.88 & 0.88 & 0.88 & 0.88 & 0.88 & 0.88 & 0.88 & 0.88 & 0.88 & 0.87 & 0.87 & 0.87 & 0.87 \\
                \midrule[1.125pt]
                \bottomrule[1.5pt]
            \end{tabular}%
        }
    \end{center}
    \footnotesize{*we set $\beta =$ $0.1$, $0.2$, $0.3$ (where $\beta$ is byzantine aggregation heuristic, for \textsc{TrimAvg} and \textsc{MultiKrum} respectively under $10\%$, $20\%$, and $30\%$ malicious clients.}
\end{table*}
\begin{table*}[!ht]
    % {\centering
    \begin{center}
        \caption{\label{tab:results-alpha-000_1} Overall test accuracy (ACC) of federated learning under benign training and poisoning attacks across malicious client fractions (\(10\%\), \(20\%\), \(30\%\)) in the extreme non-IID setting \(\alpha = 0.1\). Results are reported on MNIST, Fashion-MNIST, and CIFAR-10, comparing standard aggregation rules (\textsc{FedAvg, Median, Trimmed Mean, GeoMedian, Krum, NNM+Krum}) against our proposed GAN-based authentication variants. GAN-based defenses consistently preserve accuracy even under high adversarial presence, whereas conventional robust aggregation methods degrade significantly.}

        \resizebox{\linewidth}{!}{%
            \begin{tabular}{ll g cgcg cgcg cgcg}\toprule[1.5pt]\midrule[1.125pt]
                & \multirow{2}{*}{\textbf{Baseline}} & \multicolumn{1}{c}{\multirow{2}{*}{\parbox{1cm}{\centering No Attack}}} & \multicolumn{4}{c}{$\epsilon = 10\%$} & \multicolumn{4}{c}{$\epsilon = 20\%$} & \multicolumn{4}{c}{$\epsilon = 30\%$} \\
                \cmidrule(lr){4-7} \cmidrule(lr){8-11} \cmidrule(lr){12-15}
                &  & \multicolumn{1}{c}{} & \multicolumn{1}{c}{\textbf{RN}} & \multicolumn{1}{c}{\textbf{LF}}  & \multicolumn{1}{c}{\textbf{SF}}  & \multicolumn{1}{c}{\textbf{IPM}} & \multicolumn{1}{c}{\textbf{RN}} & \multicolumn{1}{c}{\textbf{LF}}  & \multicolumn{1}{c}{\textbf{SF}}  & \multicolumn{1}{c}{\textbf{IPM}} & \multicolumn{1}{c}{\textbf{RN}} & \multicolumn{1}{c}{\textbf{LF}}  & \multicolumn{1}{c}{\textbf{SF}}  & \multicolumn{1}{c}{\textbf{IPM}} \\
				\midrule[1.125pt]
				\multirow{8}{*}{\rotatebox{90}{\textbf{MNIST}}}
				  & \textsc{FedAvg} & 0.99 & 0.17 & 0.95 & 0.10 & 0.10 & 0.17 & 0.67 & 0.10 & 0.10 & 0.10 & 0.10 & 0.10 & 0.10 \\
				~ & \textsc{Median} & 0.98 & 0.98 & 0.98 & 0.98 & 0.98 & 0.64 & 0.98 & 0.97 & 0.10 & 0.10 & 0.96 & 0.10 & 0.10 \\
				~ & \textsc{TrimAvg} & 0.98 & 0.43 & 0.98 & 0.96 & 0.10 & 0.14 & 0.98 & 0.10 & 0.10 & 0.10 & 0.96 & 0.11 & 0.10 \\
				~ & \textsc{GeoMedian} & 0.99 & 0.99 & 0.99 & 0.99 & 0.98 & 0.64 & 0.99 & 0.97 & 0.10 & 0.10 & 0.98 & 0.11 & 0.11 \\
				~ & \textsc{MultiKrum} & 0.96 & 0.53 & 0.97 & 0.10 & 0.10 & 0.33 & 0.87 & 0.10 & 0.11 & 0.10 & 0.11 & 0.11 & 0.10 \\
				~ & \textsc{NNM+Krum} & 0.98 & 0.37 & 0.98 & 0.94 & 0.10 & 0.24 & 0.97 & 0.10 & 0.10 & 0.10 & 0.97 & 0.11 & 0.10 \\ \cmidrule(lr){2-15}
				~ & \textsc{GAN (Fixed Threshold)} & 0.62 & 0.65 & 0.64 & 0.59 & 0.58 & 0.62 & 0.65 & 0.68 & 0.64 & 0.54 & 0.79 & 0.58 & 0.57 \\
				~ & \textsc{GAN (Adaptive: Mean)} & 0.98 & 0.96 & 0.97 & 0.97 & 0.98 & 0.93 & 0.98 & 0.98 & 0.98 & 0.91 & 0.11 & 0.98 & 0.98 \\
				~ & \textsc{GAN (Clustered)} & 0.98 & 0.97 & 0.98 & 0.99 & 0.11 & 0.79 & 0.97 & 0.94 & 0.98 & 0.96 & 0.11 & 0.98 & 0.11 \\
				\midrule[1.125pt]
				\multirow{8}{*}{\rotatebox{90}{\textbf{Fashion MNIST}}}
				  & \textsc{FedAvg} & 0.85 & 0.20 & 0.66 & 0.10 & 0.10 & 0.23 & 0.10 & 0.10 & 0.10 & 0.10 & 0.10 & 0.10 & 0.10 \\
				~ & \textsc{Median} & 0.83 & 0.82 & 0.83 & 0.80 & 0.80 & 0.31 & 0.78 & 0.70 & 0.10 & 0.10 & 0.78 & 0.10 & 0.10 \\
				~ & \textsc{TrimAvg} & 0.85 & 0.28 & 0.85 & 0.10 & 0.10 & 0.38 & 0.78 & 0.10 & 0.10 & 0.10 & 0.77 & 0.10 & 0.10 \\
				~ & \textsc{GeoMedian} & 0.86 & 0.84 & 0.85 & 0.84 & 0.77 & 0.56 & 0.82 & 0.74 & 0.10 & 0.10 & 0.80 & 0.10 & 0.10 \\
				~ & \textsc{MultiKrum} & 0.83 & 0.41 & 0.81 & 0.10 & 0.10 & 0.10 & 0.79 & 0.10 & 0.10 & 0.10 & 0.74 & 0.10 & 0.10 \\
				~ & \textsc{NNM+Krum} & 0.85 & 0.51 & 0.83 & 0.10 & 0.10 & 0.40 & 0.80 & 0.10 & 0.10 & 0.10 & 0.72 & 0.10 & 0.10 \\ \cmidrule(lr){2-15}
				~ & \textsc{GAN (Fixed Threshold)} & 0.33 & 0.35 & 0.45 & 0.33 & 0.36 & 0.32 & 0.31 & 0.35 & 0.41 & 0.29 & 0.33 & 0.32 & 0.30 \\
				~ & \textsc{GAN (Adaptive: Mean)} & 0.10 & 0.10 & 0.10 & 0.10 & 0.10 & 0.55 & 0.10 & 0.10 & 0.10 & 0.10 & 0.10 & 0.10 & 0.10 \\
				~ & \textsc{GAN (Clustered)} & 0.10 & 0.62 & 0.10 & 0.10 & 0.10 & 0.48 & 0.10 & 0.10 & 0.83 & 0.10 & 0.10 & 0.10 & 0.83 \\
				\midrule[1.125pt]
				\multirow{8}{*}{\rotatebox{90}{\textbf{CIFAR-10}}}
				  & \textsc{FedAvg} & 0.69 & 0.11 & 0.62 & 0.13 & 0.10 & 0.10 & 0.54 & 0.10 & 0.10 & 0.12 & 0.33 & 0.10 & 0.10 \\
				~ & \textsc{Median} & 0.70 & 0.60 & 0.57 & 0.58 & 0.45 & 0.16 & 0.52 & 0.48 & 0.13 & 0.12 & 0.42 & 0.19 & 0.10 \\
				~ & \textsc{TrimAvg} & 0.67 & 0.11 & 0.65 & 0.53 & 0.16 & 0.10 & 0.57 & 0.45 & 0.10 & 0.11 & 0.39 & 0.13 & 0.10 \\
				~ & \textsc{GeoMedian} & 0.61 & 0.67 & 0.64 & 0.65 & 0.47 & 0.17 & 0.61 & 0.55 & 0.20 & 0.12 & 0.42 & 0.32 & 0.10 \\
				~ & \textsc{MultiKrum} & 0.69 & 0.11 & 0.66 & 0.55 & 0.10 & 0.12 & 0.58 & 0.37 & 0.10 & 0.10 & 0.53 & 0.22 & 0.10 \\
				~ & \textsc{NNM+Krum} & 0.69 & 0.11 & 0.66 & 0.50 & 0.10 & 0.11 & 0.62 & 0.28 & 0.10 & 0.10 & 0.60 & 0.25 & 0.10 \\ \cmidrule(lr){2-15}
				~ & \textsc{GAN (Fixed Threshold)} & 0.29 & 0.22 & 0.25 & 0.25 & 0.26 & 0.25 & 0.26 & 0.25 & 0.27 & 0.26 & 0.24 & 0.25 & 0.26 \\
				~ & \textsc{GAN (Adaptive: Mean)} & 0.71 & 0.66 & 0.63 & 0.67 & 0.64 & 0.70 & 0.63 & 0.67 & 0.67 & 0.67 & 0.62 & 0.41 & 0.67 \\
				~ & \textsc{GAN (Clustered)} & 0.73 & 0.60 & 0.59 & 0.58 & 0.60 & 0.68 & 0.69 & 0.52 & 0.68 & 0.68 & 0.49 & 0.24 & 0.66 \\
                \midrule[1.125pt]
                \bottomrule[1.5pt]
            \end{tabular}%
        }
    \end{center}
    \footnotesize{*we set $\beta =$ $0.1$, $0.2$, $0.3$ (where $\beta$ is byzantine aggregation heuristic, for \textsc{TrimAvg} and \textsc{MultiKrum} respectively under $10\%$, $20\%$, and $30\%$ malicious clients.}
\end{table*}
\begin{table*}[!t]
    \begin{center}
        \caption{\label{tab:tpr-alpha-100} True Positive Rate (TPR) of all evaluated defense methods across datasets and malicious client ratios, shown for Dirichlet $\alpha=100$. Higher values indicate stronger detection of malicious updates.}
        \resizebox{\linewidth}{!}{%
            \begin{tabular}{ll g cgcg cgcg cgcg}\toprule[1.5pt]\midrule[1.125pt]
                & \multirow{2}{*}{\textbf{Baseline}} & \multicolumn{1}{c}{\multirow{2}{*}{\parbox{1cm}{\centering No Attack}}} & \multicolumn{4}{c}{$\epsilon = 10\%$} & \multicolumn{4}{c}{$\epsilon = 20\%$} & \multicolumn{4}{c}{$\epsilon = 30\%$} \\
                \cmidrule(lr){4-7} \cmidrule(lr){8-11} \cmidrule(lr){12-15}
                &  & \multicolumn{1}{c}{} & \multicolumn{1}{c}{\textbf{RN}} & \multicolumn{1}{c}{\textbf{LF}}  & \multicolumn{1}{c}{\textbf{SF}}  & \multicolumn{1}{c}{\textbf{IPM}} & \multicolumn{1}{c}{\textbf{RN}} & \multicolumn{1}{c}{\textbf{LF}}  & \multicolumn{1}{c}{\textbf{SF}}  & \multicolumn{1}{c}{\textbf{IPM}} & \multicolumn{1}{c}{\textbf{RN}} & \multicolumn{1}{c}{\textbf{LF}}  & \multicolumn{1}{c}{\textbf{SF}}  & \multicolumn{1}{c}{\textbf{IPM}} \\
				\midrule[1.125pt]
				\multirow{4}{*}{\rotatebox{90}{\textbf{MNIST}}}
				~ & \textsc{MultiKrum} & 0.00 & 0.68 & 0.67 & 0.67 & 0.68 & 0.64 & 0.74 & 0.75 & 0.42 & 0.33 & 0.70 & 0.72 & 0.28 \\
				~ & \textsc{GAN (Fixed Threshold)} & 0.00 & 1.00 & 1.00 & 1.00 & 1.00 & 1.00 & 1.00 & 1.00 & 1.00 & 1.00 & 1.00 & 1.00 & 1.00 \\
				~ & \textsc{GAN (Adaptive: Mean)} & 0.00 & 1.00 & 1.00 & 1.00 & 1.00 & 1.00 & 1.00 & 1.00 & 1.00 & 1.00 & 1.00 & 1.00 & 1.00 \\
				~ & \textsc{GAN (Clustered)} & 0.00 & 1.00 & 1.00 & 1.00 & 1.00 & 1.00 & 1.00 & 1.00 & 1.00 & 1.00 & 1.00 & 1.00 & 1.00 \\
				\midrule[1.125pt]
				\multirow{4}{*}{\rotatebox{90}{\textbf{FMNIST}}}
				~ & \textsc{MultiKrum} & 0.00 & 0.68 & 0.68 & 0.60 & 0.54 & 0.73 & 0.74 & 0.75 & 0.32 & 0.33 & 0.65 & 0.63 & 0.29 \\
				~ & \textsc{GAN (Fixed Threshold)} & 0.00 & 1.00 & 1.00 & 1.00 & 1.00 & 1.00 & 1.00 & 1.00 & 1.00 & 1.00 & 1.00 & 1.00 & 1.00 \\
				~ & \textsc{GAN (Adaptive: Mean)} & 0.00 & 1.00 & 1.00 & 1.00 & 1.00 & 1.00 & 1.00 & 1.00 & 1.00 & 1.00 & 1.00 & 1.00 & 1.00 \\
				~ & \textsc{GAN (Clustered)} & 0.00 & 1.00 & 1.00 & 1.00 & 1.00 & 1.00 & 1.00 & 1.00 & 1.00 & 1.00 & 1.00 & 1.00 & 1.00 \\
				\midrule[1.125pt]
				\multirow{4}{*}{\rotatebox{90}{\textbf{CIFAR-10}}}
				~ & \textsc{MultiKrum} & 0.00 & 0.68 & 0.68 & 0.68 & 0.57 & 0.73 & 0.75 & 0.76 & 0.71 & 0.70 & 0.76 & 0.83 & 0.53 \\
				~ & \textsc{GAN (Fixed Threshold)} & 0.00 & 1.00 & 1.00 & 1.00 & 1.00 & 1.00 & 1.00 & 1.00 & 1.00 & 1.00 & 1.00 & 1.00 & 1.00 \\
				~ & \textsc{GAN (Adaptive: Mean)} & 0.00 & 1.00 & 1.00 & 1.00 & 1.00 & 1.00 & 1.00 & 1.00 & 1.00 & 1.00 & 1.00 & 1.00 & 1.00 \\
				~ & \textsc{GAN (Clustered)} & 0.00 & 1.00 & 1.00 & 1.00 & 1.00 & 1.00 & 1.00 & 1.00 & 1.00 & 1.00 & 1.00 & 1.00 & 1.00 \\
                \midrule[1.125pt]
                \bottomrule[1.5pt]
            \end{tabular}%
        }
    \end{center}
\end{table*}

\begin{table*}[!t]
    \begin{center}
        \caption{\label{tab:tnr-alpha-100} True Negative Rate (TNR) of all evaluated defense methods across datasets and malicious client ratios, shown for Dirichlet $\alpha=100$. Higher values indicate better preservation of benign updates.}
        \resizebox{\linewidth}{!}{%
            \begin{tabular}{ll g cgcg cgcg cgcg}\toprule[1.5pt]\midrule[1.125pt]
                & \multirow{2}{*}{\textbf{Baseline}} & \multicolumn{1}{c}{\multirow{2}{*}{\parbox{1cm}{\centering No Attack}}} & \multicolumn{4}{c}{$\epsilon = 10\%$} & \multicolumn{4}{c}{$\epsilon = 20\%$} & \multicolumn{4}{c}{$\epsilon = 30\%$} \\
                \cmidrule(lr){4-7} \cmidrule(lr){8-11} \cmidrule(lr){12-15}
                &  & \multicolumn{1}{c}{} & \multicolumn{1}{c}{\textbf{RN}} & \multicolumn{1}{c}{\textbf{LF}}  & \multicolumn{1}{c}{\textbf{SF}}  & \multicolumn{1}{c}{\textbf{IPM}} & \multicolumn{1}{c}{\textbf{RN}} & \multicolumn{1}{c}{\textbf{LF}}  & \multicolumn{1}{c}{\textbf{SF}}  & \multicolumn{1}{c}{\textbf{IPM}} & \multicolumn{1}{c}{\textbf{RN}} & \multicolumn{1}{c}{\textbf{LF}}  & \multicolumn{1}{c}{\textbf{SF}}  & \multicolumn{1}{c}{\textbf{IPM}} \\
				\midrule[1.125pt]
				\multirow{4}{*}{\rotatebox{90}{\textbf{MNIST}}}
				~ & \textsc{MultiKrum} & 1.00 & 0.96 & 0.96 & 0.96 & 0.96 & 0.90 & 0.92 & 0.93 & 0.85 & 0.71 & 0.86 & 0.87 & 0.69 \\
				~ & \textsc{GAN (Fixed Threshold)} & 0.99 & 1.00 & 0.99 & 1.00 & 0.88 & 0.99 & 0.98 & 1.00 & 1.00 & 0.99 & 0.98 & 1.00 & 0.99 \\
				~ & \textsc{GAN (Adaptive: Mean)} & 0.79 & 0.91 & 0.91 & 0.91 & 0.91 & 0.95 & 0.95 & 0.95 & 0.95 & 0.98 & 0.98 & 0.98 & 0.98 \\
				~ & \textsc{GAN (Clustered)} & 0.85 & 0.93 & 0.94 & 0.94 & 0.93 & 0.97 & 0.97 & 0.97 & 0.97 & 0.98 & 0.98 & 0.98 & 0.98 \\
				\midrule[1.125pt]
				\multirow{4}{*}{\rotatebox{90}{\textbf{FMNIST}}}
				~ & \textsc{MultiKrum} & 1.00 & 0.96 & 0.96 & 0.95 & 0.95 & 0.92 & 0.92 & 0.93 & 0.83 & 0.71 & 0.84 & 0.83 & 0.70 \\
				~ & \textsc{GAN (Fixed Threshold)} & 0.81 & 0.79 & 0.79 & 0.77 & 0.79 & 0.79 & 0.48 & 0.79 & 0.78 & 0.82 & 0.80 & 0.80 & 0.80 \\
				~ & \textsc{GAN (Adaptive: Mean)} & 0.74 & 0.79 & 0.81 & 0.78 & 0.78 & 0.81 & 0.83 & 0.81 & 0.80 & 0.85 & 0.84 & 0.82 & 0.84 \\
				~ & \textsc{GAN (Clustered)} & 0.78 & 0.80 & 0.83 & 0.80 & 0.81 & 0.84 & 0.85 & 0.85 & 0.88 & 0.87 & 0.88 & 0.87 & 0.86 \\
				\midrule[1.125pt]
				\multirow{4}{*}{\rotatebox{90}{\textbf{CIFAR-10}}}
				~ & \textsc{MultiKrum} & 1.00 & 0.96 & 0.96 & 0.96 & 0.95 & 0.92 & 0.93 & 0.93 & 0.92 & 0.86 & 0.88 & 0.91 & 0.79 \\
				~ & \textsc{GAN (Fixed Threshold)} & 0.03 & 0.03 & 0.03 & 0.03 & 0.03 & 0.03 & 0.03 & 0.03 & 0.03 & 0.03 & 0.03 & 0.03 & 0.03 \\
				~ & \textsc{GAN (Adaptive: Mean)} & 0.60 & 0.83 & 0.83 & 0.83 & 0.83 & 0.92 & 0.91 & 0.92 & 0.92 & 0.96 & 0.96 & 0.96 & 0.96 \\
				~ & \textsc{GAN (Clustered)} & 0.68 & 0.87 & 0.86 & 0.87 & 0.86 & 0.93 & 0.93 & 0.93 & 0.94 & 0.97 & 0.96 & 0.96 & 0.96 \\
                \midrule[1.125pt]
                \bottomrule[1.5pt]
            \end{tabular}%
        }
    \end{center}
\end{table*}
% \input{figures/00-Fig4-Ablation-Convergence}
% \input{tables/MAIN-RES-TABLE-02-010_0}
% \input{tables/MAIN-RES-TABLE-03-001_0}
%%%%%%%%%%%%%%%%%%%%%%%%%%%%%%%%%%%%%%%%%%%%%%%%%%%%%%%%%%%%%%%%%%%%%%%%%%%%%%%%

\section{Experimental Results}
\noindent 
We compare our method against state-of-the-art baselines under various adversarial settings and analyze its performance using the evaluation metrics described in Section~\ref{sec:eval_metrics}. Results across datasets and poisoning scenarios consistently show that our GAN-based defenses achieve both high task accuracy and strong detection performance, outperforming standard aggregation rules.

%%%%%%%%%%%%%%%%%%%%%%%%%%%%%%%%%%%%%%%%%%%%%%%%%%%%%%%%%%%%%%%%%%%%%%%%%%%%%%%%
\begin{figure*}[!t]
    \centering
    \includegraphics[page=1, width=1.0\linewidth]{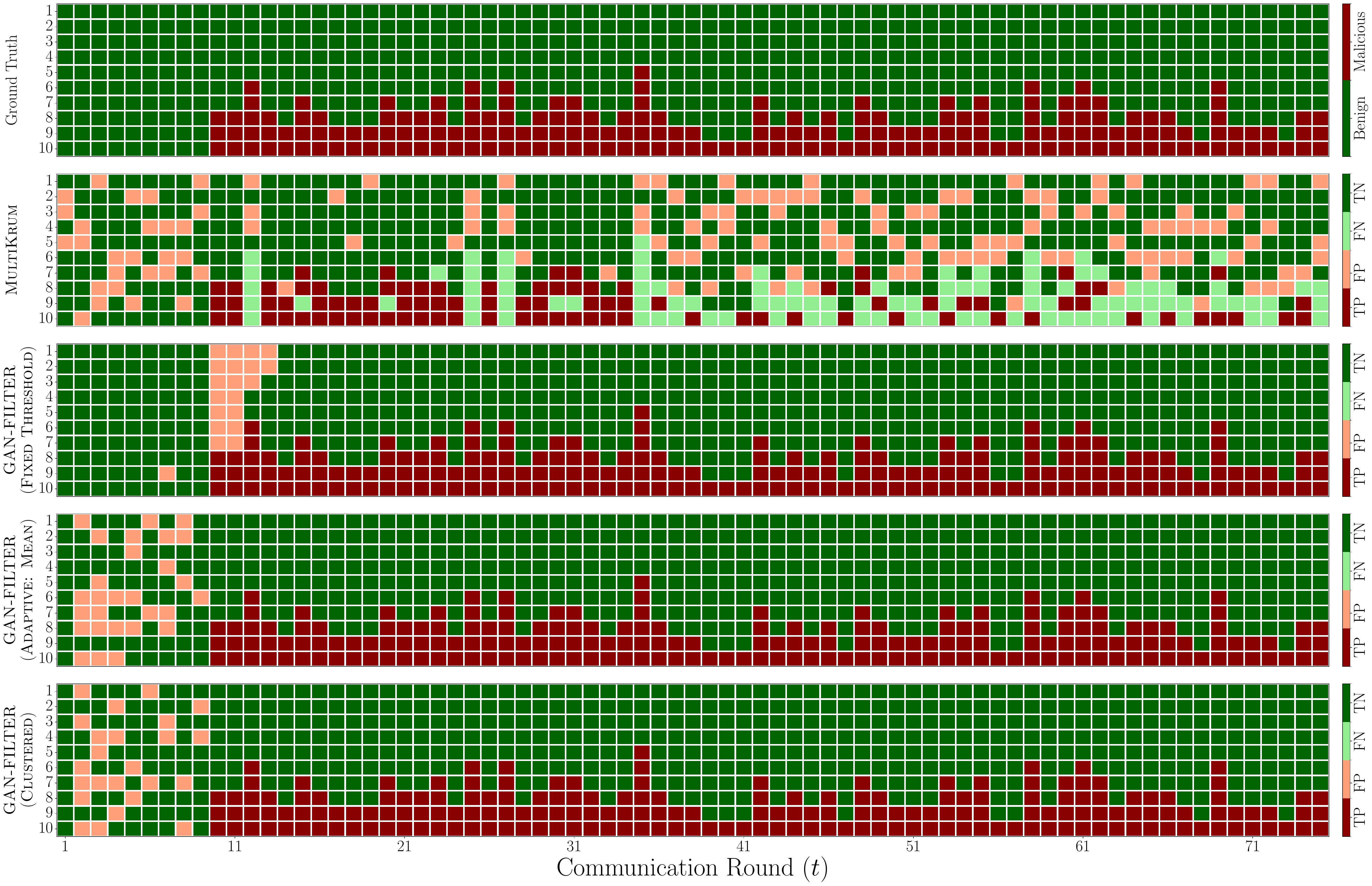}
    \caption{Heatmap of detection performance for \textsc{MultiKrum} and proposed variants under a Random Noise attack ($30\%$ malicious clients) on the MNIST dataset. The proposed GAN-based framework achieves high TPR and TNR, outperforming \textsc{MultiKrum} in detecting malicious updates.}
    \label{fig:heatmap_mnist}
\end{figure*}
%%%%%%%%%%%%%%%%%%%%%%%%%%%%%%%%%%%%%%%%%%%%%%%%%%%%%%%%%%%%%%%%%%%%%%%%%%%%%%%%

\subsection{Robustness Against Poisoning Attacks}
\noindent
We evaluate robustness against the four poisoning attacks from Section~\ref{sec:attack_setup}. Across datasets, conventional methods (\textsc{FedAvg}, \textsc{Median}, \textsc{TrimAvg}, \textsc{GeoMedian}, \textsc{MultiKrum}, \textsc{NNM+Krum}) perform well without adversaries but degrade sharply once malicious clients are introduced. For instance, under Sign Flipping with $30\%$ adversaries on MNIST, \textsc{FedAvg} and \textsc{Krum} collapse to $10\%$ accuracy, while our adaptive GAN variants sustain $99\%$. Robust baselines such as \textsc{TrimAvg} and \textsc{GeoMedian} resist simple attacks but fail under stronger ones (e.g., IPM). In contrast, GAN-based defenses consistently maintain high ACC while neutralizing diverse poisoning strategies.

\subsection{Impact of Statistical Heterogeneity}
To study statistical heterogeneity, we partition data using a class-wise Dirichlet distribution with concentration parameter $\alpha$. We evaluate four regimes: near-IID ($\alpha=100$), low heterogeneity ($\alpha=10$), moderate heterogeneity ($\alpha=1$), and extreme heterogeneity ($\alpha=0.1$).  

Table~\ref{tab:results-alpha-100} (near-IID) and Table~\ref{tab:results-alpha-000_1} (extreme non-IID) capture the extremes. In the near-IID case, all methods start with high baseline accuracy, yet standard defenses degrade rapidly as the adversarial fraction grows. In the extreme case ($\alpha=0.1$), degradation is severe: \textsc{FedAvg} and \textsc{Krum} collapse, and robust baselines lose effectiveness. By contrast, our adaptive GANs preserve high accuracy even at $\epsilon=30\%$, demonstrating resilience to both non-IID data and adversarial pressure. Notably, under benign IID conditions, the fixed-threshold variant achieves near-perfect performance (e.g., ACC=$0.99$ on MNIST with $30\%$ malicious clients; Table~\ref{tab:results-alpha-100}), effectively mitigating Random Noise, Sign Flipping, Label-Flipping, and IPM attacks.  

\subsection{Intermediate Settings}
In low and moderate non-IID regimes (Table~\ref{tab:results-alpha-010} and Table~\ref{tab:results-alpha-001} respectively), trends mirror those at the extremes: baseline methods rapidly degrade as heterogeneity and adversarial proportion increase, while GAN-based defenses remain stable. For instance, with $\alpha=10$ and $\epsilon=30\%$, \textsc{FedAvg} accuracy drops to $10\%$, and \textsc{Krum} also collapses to $10\%$, whereas our adaptive GAN maintains $99\%$. Similarly, at $\alpha=1.0$, the fixed-threshold GAN sustains above $99\%$ accuracy even under $30\%$ adversaries, while conventional defenses fall below $10\%$. Among our variants, the fixed-threshold GAN is conservative but steady, whereas the adaptive and clustered GANs achieve the best balance of accuracy and adaptability, consistently outperforming existing baselines across both intermediate settings.

\subsection{Performance Under Benign Settings}
To ensure defenses do not degrade normal learning, we evaluate ACC in the absence of malicious clients. On MNIST, Fashion-MNIST, and CIFAR-10, our methods achieve competitive accuracy under both IID ($\alpha=100.0$) and non-IID ($\alpha=0.1$) conditions (first column of Tables~\ref{tab:results-alpha-100}--\ref{tab:results-alpha-001}), confirming that robustness is not attained at the expense of benign performance.

\subsection{Detection Performance}
Beyond accuracy, we assess the detection ability of our framework in terms of True Positive Rate (TPR) and True Negative Rate (TNR), which measure how effectively malicious updates are identified while preserving benign contributions. We compare against \textsc{MultiKrum}, a widely used Byzantine-resilient baseline that also performs selective aggregation. 

Our adaptive-threshold variants achieve near-perfect detection of malicious clients. On MNIST with $10\%$ malicious clients, the adaptive method maintains a TPR between $0.98$--$1.00$ across all attack types, compared to $0.54$--$0.68$ for \textsc{MultiKrum}. Even under stronger adversarial pressure ($30\%$ malicious clients), and extreme heterogeneity ($\alpha=0.1$) the adaptive GAN achieves a TPR of $0.88$, outperforming \textsc{MultiKrum} (TPR = $0.31$). Similar patterns hold across datasets: on CIFAR-10 with $20\%$ malicious clients, the adaptive GAN sustains TPR $\approx 0.97$ while \textsc{MultiKrum} drops below $0.20$. Detailed TPR results for all datasets, malicious ratios, and Dirichlet distributions are provided in Tables~\ref{tab:tpr-alpha-100}, \ref{tab:tpr-alpha-010}, \ref{tab:tpr-alpha-001}, \ref{tab:tpr-alpha-000_1} in the Appendix~\ref{sec:appendix-b}. Note that the True Positive Rate (TPR) is 0 in the absence of malicious clients, since there are no positive cases to detect.

The framework also preserves benign updates with high accuracy. On Fashion-MNIST with $30\%$ malicious clients, adaptive and clustered variants achieve TNR in the range of $0.82$–$0.85$, compared to $0.70$–$0.84$ for \textsc{MultiKrum}. Full TNR results are reported in Tables~\ref{tab:tnr-alpha-100}, \ref{tab:tnr-alpha-010}, \ref{tab:tnr-alpha-001}, \ref{tab:tnr-alpha-000_1} in the Appendix~\ref{sec:appendix-b}. Even in non-IID regimes, both TPR and TNR remain high, underscoring robustness to statistical heterogeneity.

\textbf{Thresholding strategies.} We observe that fixed-threshold filtering is less reliable, particularly on more complex datasets such as CIFAR-10. The difficulty lies in determining an appropriate universal threshold: what works for one dataset or attack setting often under- or over-filters in another. In contrast, adaptive and clustering strategies automatically adjust to the distribution of client performance, yielding more consistent detection. While our framework produces the synthetic dataset used for evaluation, the choice of thresholding strategy ultimately lies with the deployer depending on system requirements and operational context.

To further illustrate these trends, the MNIST TPR/TNR heatmap under Random Noise attacks with $30\%$ malicious clients is shown in Figure~\ref{fig:heatmap_mnist}, while heatmaps for Fashion-MNIST and CIFAR-10 are provided in Appendix~\ref{sec:appendix-b}. For clarity, only the first $75$ communication rounds (out of $200$) are displayed. Across datasets, our GAN-based defenses consistently outperform \textsc{MultiKrum}, achieving higher detection rates for adversarial updates while better preserving benign contributions.

A subtle trade-off arises in adversary-free scenarios: the adaptive strategy, by design, always filters a fraction of updates, which slightly lowers TNR due to increased false positives. However, fixed-threshold and clustering-based variants are less affected, and accuracy remains competitive. This highlights a tunable balance between conservativeness (fixed) and adaptability (adaptive), depending on deployment requirements.

%%%%%%%%%%%%%%%%%%%%%%%%%%%%%%%%%%%%%%%%%%%%%%%%%%%%%%%%%%%%%%%%%%%%%%%%%%%%%%%%
\begin{figure}[!t]
    \centering
    \includegraphics[width=\linewidth]{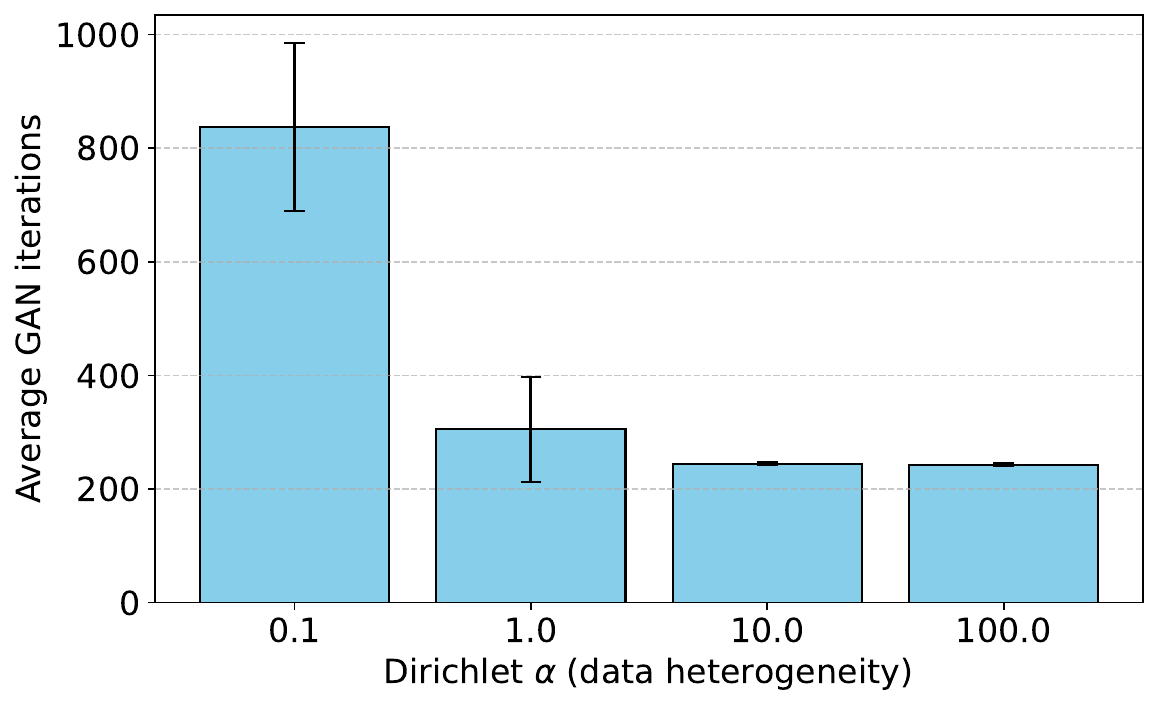}
    % \caption{Average number of GAN training iterations until early stopping across different Dirichlet $\alpha$ values. Smaller $\alpha$ corresponds to higher non-IID heterogeneity among clients. We observe that IID settings (large $\alpha$) lead to a sharper reduction in training iterations, while non-IID settings (small $\alpha$) retain higher overhead due to greater update variance.}
    \caption{Average number of GAN training iterations required for the generator to reach early stopping across different Dirichlet $\alpha$ values. Smaller $\alpha$ corresponds to higher non-IID heterogeneity among clients. The plot shows that as $\alpha$ decreases (more heterogeneous data), the GAN requires more iterations on average, while IID settings (larger $\alpha$) lead to faster convergence.}
    \label{fig:gan_iterations_alpha}
\end{figure}

\begin{figure}[!t]
    \centering
    \includegraphics[width=\linewidth]{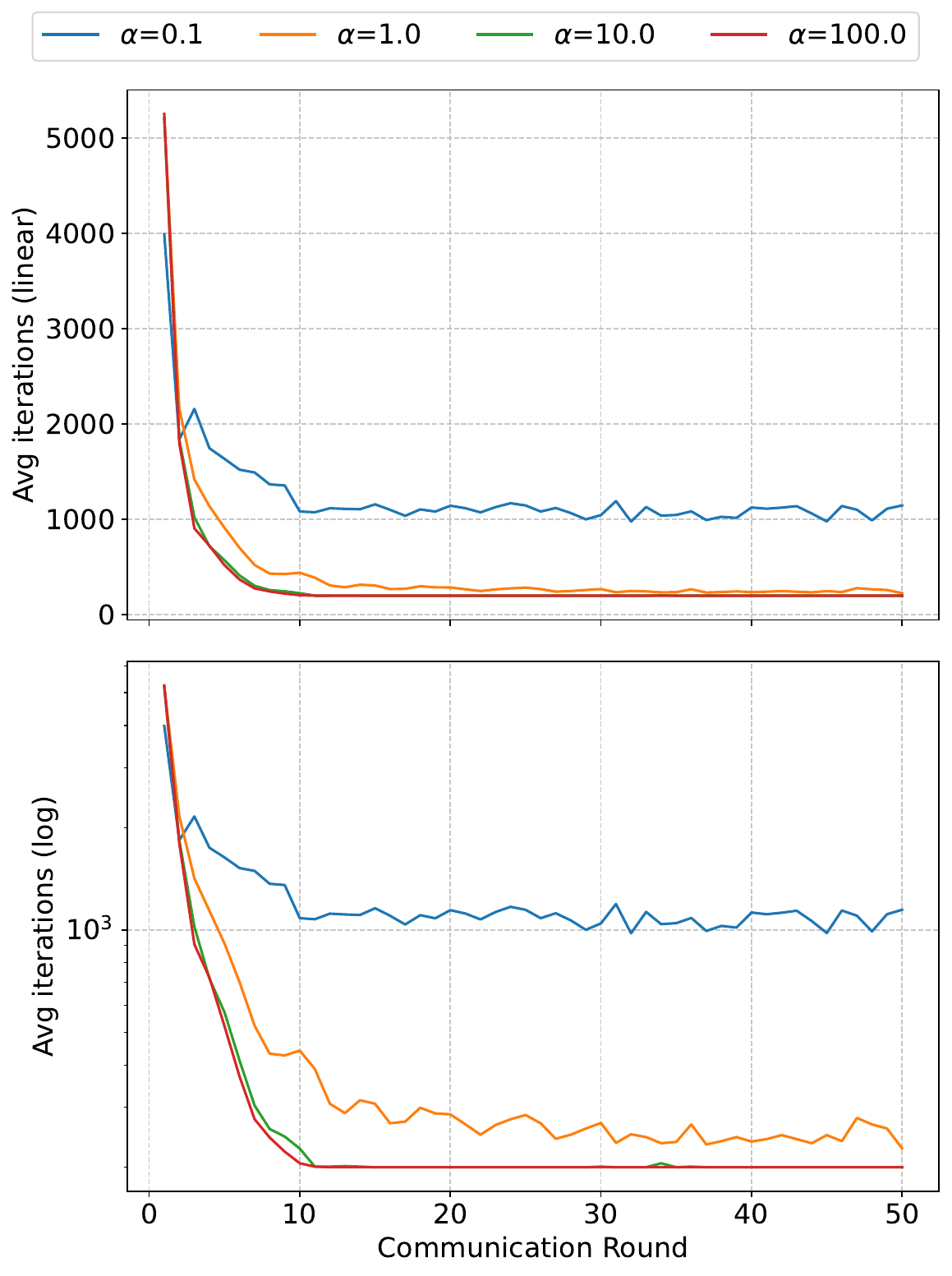}
    \caption{Evolution of GAN training iterations per communication round for different Dirichlet $\alpha$ values on the CIFAR-10 dataset (first 50 rounds shown for clarity).. The top subplot (linear scale) highlights absolute iteration trends, while the bottom subplot (log scale) emphasizes relative changes over rounds. Each line shows how the number of iterations required for the generator to converge decreases as FL training progresses. Smaller $\alpha$ values (higher non-IID heterogeneity) require more iterations initially and decline more gradually, whereas larger $\alpha$ (IID) settings converge faster.}
    \label{fig:gan_iterations_rounds}
\end{figure}

%%%%%%%%%%%%%%%%%%%%%%%%%%%%%%%%%%%%%%%%%%%%%%%%%%%%%%%%%%%%%%%%%%%%%%%%%%%%%%%%

% \subsection{Impact of Convergence} 
% Finally, we analyze the effect of initial model convergence on defense. We pre-train the global model for $K\in\{5,10,15,20\}$ benign rounds before introducing IPM attacks on CIFAR-10. Results in Figure~\ref{fig:impact_convergence} show that higher $K$ leads to better test accuracy. e.g., at $K=100$, adaptive-median and clustered GANs achieve ASR=$0.20$, compared to $0.60$ for $K=5$. This improvement stems from the global model’s stability, which enhances synthetic data generation and malicious update detection.

% The results, presented in Figure~\ref{fig:impact_convergence}, demonstrate that as \(K\) increases, the detection rate of malicious updates improves significantly, leading to a decrease in ASR for the Backdoor attack. For example, with \(K = 100\), the adaptive threshold (median) and clustered methods reduce the ASR to \(0.20\), compared to \(0.60\) for \(K = 5\) and \(K = 10\). This improvement is attributed to the global model’s increased reliability and convergence after \(K\) rounds of benign training, providing more accurate information to the cGAN for synthetic data generation and update validation.

\subsection{Training Efficiency of the GAN Defense} 

We analyze the evolution of GAN training overhead across communication rounds. The number of iterations required for stable generator convergence is highest during the early rounds of FL training and decreases sharply as the global model stabilizes. Early in training, the global model parameters are far from convergence, resulting in larger changes per round and requiring more effort from the framework to learn the evolving decision boundaries. As the model approaches stability, the GAN converges in fewer iterations. This reduction is more pronounced in IID settings, where client updates quickly cluster around the global model trajectory, whereas non-IID settings exhibit a slower decline due to higher variance and overlapping distributions. Overall, the amortized cost of deploying the GAN defense diminishes over time, making it practical for long-horizon FL training.

To quantify this effect, we group experiments by their Dirichlet $\alpha$ parameter and report the average number of GAN training iterations until convergence (Figure~\ref{fig:gan_iterations_alpha}). The results confirm that overhead is significantly lower in IID regimes (large $\alpha$), while non-IID cases (small $\alpha$) consistently require more iterations due to heterogeneous updates.

We further examine GAN dynamics across communication rounds on the CIFAR-10 dataset for representative $\alpha$ values (Figure~\ref{fig:gan_iterations_rounds}). The number of iterations needed to reach early stopping decreases steadily as FL training progresses. Near-IID settings ($\alpha = 100$) show a sharp decline, while non-IID settings (smaller $\alpha$) maintain higher iterations for longer. These trends highlight that the amortized GAN training cost diminishes over time, reinforcing the practicality of our defense in long-horizon FL.

\subsection{Summary of Results}
In summary, our experiments demonstrate that across datasets, attack types, heterogeneity levels, and adversarial proportions, the proposed GAN-based defenses consistently maintain high test accuracy (ACC) while achieving strong detection performance (TPR/TNR). In contrast, baseline methods are brittle, effective only under narrow conditions and prone to collapse under adversarial pressure. Our framework delivers robustness in both benign and adversarial scenarios, making it broadly applicable to real-world federated learning deployments.

\section{Limitations and Future Work}
\label{sec:discussion}
While our study demonstrates the effectiveness of the proposed defense framework against generic poisoning attacks, it does not cover all forms of adversarial behavior. Notably, we do not investigate \emph{targeted attacks}, such as backdoor or trigger-based poisoning, where the adversary’s goal is to embed specific malicious behaviors that remain dormant until activated. These attacks can evade detection by maintaining high accuracy on benign data while introducing subtle but harmful effects.

Our evaluation also assumes synchronous training rounds and reliable client participation. Real-world federated learning deployments often involve asynchronous updates and resource-constrained clients, which could affect the applicability of the defense.

Future work includes extending the framework to counter targeted attacks, incorporating backdoor-resilient mechanisms such as trigger detection or adversarial example analysis, and evaluating performance against adaptive adversaries. We also aim to explore large-scale, heterogeneous, and real-world deployments to test practical robustness. Additionally, investigating advanced conditional GAN architectures, such as ccGAN, could improve the quality of synthetic validation data, though at the cost of increased computational resources.
\section{Conclusion}
In this work, we proposed a GAN-based defense framework for robust federated learning against poisoning attacks. By leveraging a synthetic validation dataset generated via a conditional GAN, our approach effectively identifies and filters malicious client updates while preserving benign contributions. Extensive experiments across MNIST, Fashion-MNIST, and CIFAR-10 demonstrate that our adaptive and clustering-based filtering strategies consistently outperform conventional Byzantine-resilient aggregation methods such as \textsc{MultiKrum}, achieving higher True Positive Rates in detecting malicious updates and maintaining strong model accuracy in IID and non-IID scenarios. Our analysis also highlights the practical benefits of adaptive filtering: it requires minimal prior knowledge of the dataset and adapts to different data distributions, whereas fixed-threshold approaches struggle to generalize across heterogeneous datasets. Overall, the proposed framework provides a flexible, scalable, and effective defense mechanism for federated learning, offering both strong robustness to a wide range of poisoning strategies and practical applicability in realistic deployment scenarios.
\section*{Acknowledgments}
The distributed experiments in this work were enabled by access to high-performance computing resources provided by a national academic supercomputing infrastructure. Additionally, the authors acknowledge the use of an AI language models (DeepSeek-V3 and ChatGPT 4.0) for assistance in refining the manuscript, including improving clarity, structure, and language. The final content and intellectual contributions remain the sole responsibility of the authors.

% The distributed experiments were enabled by resources provided by the National Academic Infrastructure for Supercomputing in Sweden (NAISS), partially funded by the Swedish Research Council through grant agreement no. 2022-06725. Additionally, the authors acknowledge the use of DeepSeek-V3, an AI language model, for assistance in refining the manuscript, including improving clarity, structure, and language. The final content and intellectual contributions remain the sole responsibility of the authors.
% optional clearing of the page
% \cleardoublepage
\appendix
\section*{Ethical Considerations}
This work studies poisoning attacks and defenses in federated learning (FL). While developing attack strategies such as label flipping and inner-product manipulation is necessary to evaluate the robustness of defenses, it raises ethical concerns since these methods could, in principle, be misused by adversaries. To mitigate this risk, all experiments are restricted to standard benchmark datasets (e.g., MNIST, CIFAR-10) that do not involve personal or sensitive information, and the contributions are framed primarily around defense evaluation.

In line with responsible disclosure and reproducibility practices, we do not release source code during the anonymous review process. A full release will be made publicly available upon acceptance, ensuring that dissemination occurs within the proper scientific context and supports constructive use.

The value of this work lies in exposing vulnerabilities to strengthen the security and trustworthiness of FL deployments, not in enabling misuse. Consistent with the USENIX ethics policy, we commit to transparency of limitations, careful contextualization of our results, and responsible dissemination to minimize potential harms while advancing defenses for the broader community.

\section*{Open Science}
In line with the USENIX Security Open Science Policy, we commit to making all research artifacts from this work openly available upon acceptance. These will include the complete source code for model definitions, training pipelines, attack and defense implementations, as well as parameter configurations and experimental logs.

To preserve anonymity during the review process, we do not provide artifact access at this stage. Upon acceptance, however, we will release a permanent DOI-linked repository to ensure long-term availability and reproducibility. Since our experiments rely exclusively on public benchmark datasets and do not involve proprietary or sensitive data, there are no restrictions on artifact sharing.

By releasing all code and data after acceptance, we aim to maximize transparency, reproducibility, and research utility for the security and machine learning communities, while balancing responsible disclosure with the requirements of anonymous review.
% optional clearing of the page
\cleardoublepage
\bibliographystyle{plain}
\bibliography{bib-files/references-custom, bib-files/references-zotero-2, bib-files/references-zotero-3}
\cleardoublepage
\section{Detailed Experimental Setup}
\label{section:appendix-a}

\subsection{Training Hyperparameters}
All clients use SGD with Nesterov momentum (0.9) and weight decay $1\times 10^{-4}$. The learning rate is fixed at $0.01$. Benign clients train for $E=10$ local epochs with batch size $B=128$. Malicious-client training specifics are attack–dependent (below).

\subsection{Attack Details}
We evaluate the robustness of our framework against four types of poisoning attacks: Random Noise, Sign Flipping, Label Flipping, and Backdoor attacks. Below, we provide details on the implementation of each attack:

\subsubsection{\textbf{Random Noise (RN) Attack}}
Before sending updates, each malicious client perturbs its local model/gradient by additive Gaussian noise:
$$
\Delta w \leftarrow \Delta w + \varepsilon,\quad \varepsilon \sim \mathcal{N}(0,\,0.5^2 I).
$$
To maximize impact, attackers perform collusion and submit the same perturbed update (sybil-style).

\subsubsection{\textbf{Sign Flipping (SF) Attack}}
This attack destabilizes the global model by reversing and scaling the direction of gradient updates. Malicious clients reverse and scale their updates to push the global step in the wrong direction. Specifically, each malicious clients submit the following:
$$
\Delta w' = -\gamma \,\Delta w,\quad \text{with }\;\gamma=5.0.
$$

\subsubsection{\textbf{Label Flipping (LF) Attack}}
During local training, malicious clients flip the labels of a subset of training samples (e.g., changing label \(0\) to \(9\) in MNIST). This attack aims to degrade the global model's accuracy by introducing incorrect label information. In our experiments, we rotate all labels (e.g., \(0 \rightarrow 1\), \(1 \rightarrow 2\), and so on). To amplify impact, malicious clients also scale their submitted updates by a factor of \(\gamma = 4.0\).

\subsubsection{\textbf{Inner Product Manipulation (IPM) Attack}}
The core idea of the IPM attack is to push the aggregated model update in a direction that is \emph{negatively aligned} with the average benign update, thereby obstructing descent of the training loss. Consider a training round \(t\) with \(n\) total clients, among which \(m\) are adversarial and \(n-m\) are honest. Denote the local update of client \(i\) as \(w_t^{i}\), and let the average update of the honest group be
\[
\bar{w}_t^{\mathcal{H}} \;=\; \frac{1}{n-m} \sum_{i \in \mathcal{H}_t} w_t^{i}.
\]
A canonical IPM construction lets each adversarial client submit a scaled negative copy of this benign mean:
\begin{equation}
w_t^{j} = -\gamma\,\bar{w}^{\mathcal{H}}_t, \qquad \forall j \in \mathcal{B}_t \label{eq:ipm-payload}
\end{equation}
where \(\gamma>0\) is a tunable attack parameter that determines how aggressively the adversary amplifies the opposition. In practice, the exact value of \(\gamma\) that guarantees maximal disruption is unknown. To address this, each malicious client performs a one-dimensional line search over \(\gamma\). The idea is to probe multiple candidate scalings of the form
\[
w_t^{j}(\gamma) = -\gamma\,\tilde{w}_t, 
\]
where \(\tilde{w}_t\) is the adversary’s estimate of the benign mean (e.g., obtained from its own local update or a running historical approximation). For each candidate \(\gamma\), the adversary evaluates a surrogate loss on its proxy dataset and selects the value of \(\gamma\) that maximizes the inner product
\[
\langle w_t^{\,j}(\gamma),\, \tilde{w}_t \rangle,
\]
thereby ensuring strongest opposition to the benign direction.

\begin{figure*}[!t]
    \centering
    \includegraphics[page=1, width=1.0\linewidth]{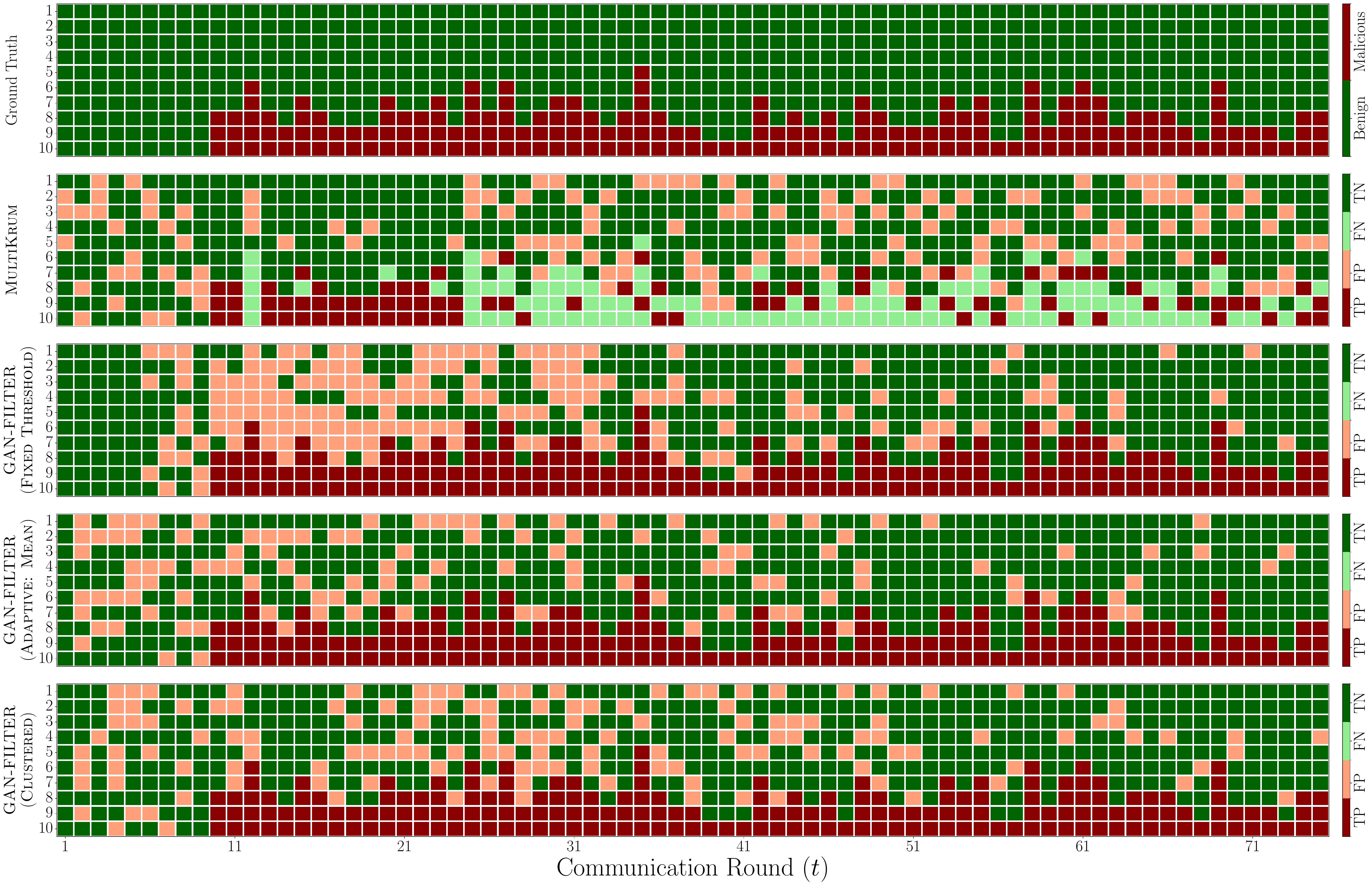}
    \caption{Heatmap comparing \textsc{MultiKrum} and proposed method variants under a Random Noise attack under a Random Noise attack with $30\%$ malicious clients on the Fashion-MNIST dataset. }
    \label{fig:heatmap_fmnist}
\end{figure*}
\begin{figure*}[!t]
    \centering
    \includegraphics[page=1, width=1.0\linewidth]{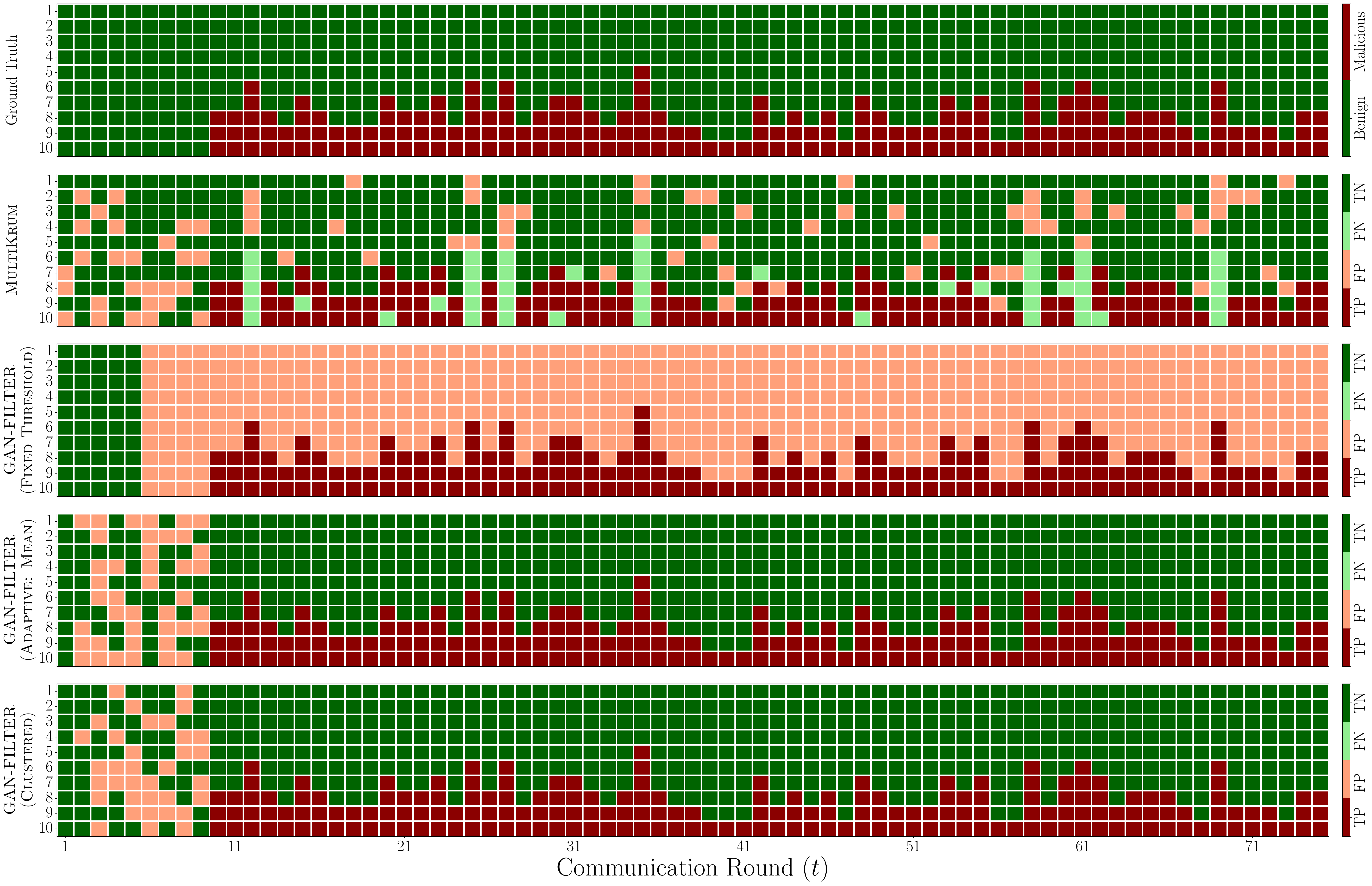}
    \caption{Heatmap showing \textsc{MultiKrum} and proposed method performance under a Random Noise attack ($30\%$ malicious clients) on the CIFAR-10 dataset. The proposed framework demonstrates robust detection capabilities, outperforming \textsc{MultiKrum} in complex settings.}
    \label{fig:heatmap_cifar10}
\end{figure*}

\begin{table*}[!t]
    \begin{center}
        \caption{\label{tab:tpr-alpha-010} True Positive Rate (TPR) of all evaluated defense methods across datasets and malicious client ratios, shown for Dirichlet $\alpha=10$. Higher values indicate stronger detection of malicious updates.}
        \resizebox{\linewidth}{!}{%
            \begin{tabular}{ll g cgcg cgcg cgcg}\toprule[1.5pt]\midrule[1.125pt]
                & \multirow{2}{*}{\textbf{Baseline}} & \multicolumn{1}{c}{\multirow{2}{*}{\parbox{1cm}{\centering No Attack}}} & \multicolumn{4}{c}{$\epsilon = 10\%$} & \multicolumn{4}{c}{$\epsilon = 20\%$} & \multicolumn{4}{c}{$\epsilon = 30\%$} \\
                \cmidrule(lr){4-7} \cmidrule(lr){8-11} \cmidrule(lr){12-15}
                &  & \multicolumn{1}{c}{} & \multicolumn{1}{c}{\textbf{RN}} & \multicolumn{1}{c}{\textbf{LF}}  & \multicolumn{1}{c}{\textbf{SF}}  & \multicolumn{1}{c}{\textbf{IPM}} & \multicolumn{1}{c}{\textbf{RN}} & \multicolumn{1}{c}{\textbf{LF}}  & \multicolumn{1}{c}{\textbf{SF}}  & \multicolumn{1}{c}{\textbf{IPM}} & \multicolumn{1}{c}{\textbf{RN}} & \multicolumn{1}{c}{\textbf{LF}}  & \multicolumn{1}{c}{\textbf{SF}}  & \multicolumn{1}{c}{\textbf{IPM}} \\
				\midrule[1.125pt]
				\multirow{4}{*}{\rotatebox{90}{\textbf{MNIST}}}
				~ & \textsc{MultiKrum} & 0.00 & 0.68 & 0.67 & 0.54 & 0.61 & 0.64 & 0.68 & 0.68 & 0.39 & 0.33 & 0.63 & 0.77 & 0.30 \\
				~ & \textsc{GAN (Fixed Threshold)} & 0.00 & 1.00 & 1.00 & 1.00 & 1.00 & 1.00 & 1.00 & 1.00 & 1.00 & 1.00 & 1.00 & 1.00 & 1.00 \\
				~ & \textsc{GAN (Adaptive: Mean)} & 0.00 & 1.00 & 1.00 & 1.00 & 1.00 & 1.00 & 1.00 & 1.00 & 1.00 & 1.00 & 1.00 & 1.00 & 1.00 \\
				~ & \textsc{GAN (Clustered)} & 0.00 & 1.00 & 1.00 & 1.00 & 1.00 & 1.00 & 1.00 & 1.00 & 1.00 & 1.00 & 1.00 & 1.00 & 1.00 \\
				\midrule[1.125pt]
				\multirow{4}{*}{\rotatebox{90}{\textbf{FMNIST}}}
				~ & \textsc{MultiKrum} & 0.00 & 0.68 & 0.67 & 0.49 & 0.56 & 0.73 & 0.73 & 0.59 & 0.30 & 0.34 & 0.59 & 0.72 & 0.29 \\
				~ & \textsc{GAN (Fixed Threshold)} & 0.00 & 1.00 & 1.00 & 1.00 & 1.00 & 1.00 & 1.00 & 1.00 & 1.00 & 1.00 & 1.00 & 1.00 & 1.00 \\
				~ & \textsc{GAN (Adaptive: Mean)} & 0.00 & 1.00 & 1.00 & 1.00 & 1.00 & 1.00 & 1.00 & 1.00 & 1.00 & 1.00 & 1.00 & 1.00 & 1.00 \\
				~ & \textsc{GAN (Clustered)} & 0.00 & 1.00 & 1.00 & 1.00 & 1.00 & 1.00 & 1.00 & 1.00 & 1.00 & 1.00 & 1.00 & 1.00 & 1.00 \\
				\midrule[1.125pt]
				\multirow{4}{*}{\rotatebox{90}{\textbf{CIFAR-10}}}
				~ & \textsc{MultiKrum} & 0.00 & 0.68 & 0.68 & 0.68 & 0.57 & 0.73 & 0.75 & 0.76 & 0.70 & 0.70 & 0.77 & 0.83 & 0.54 \\
				~ & \textsc{GAN (Fixed Threshold)} & 0.00 & 1.00 & 1.00 & 1.00 & 1.00 & 1.00 & 1.00 & 1.00 & 1.00 & 1.00 & 1.00 & 1.00 & 1.00 \\
				~ & \textsc{GAN (Adaptive: Mean)} & 0.00 & 1.00 & 1.00 & 1.00 & 1.00 & 1.00 & 1.00 & 1.00 & 1.00 & 1.00 & 1.00 & 1.00 & 1.00 \\
				~ & \textsc{GAN (Clustered)} & 0.00 & 1.00 & 1.00 & 1.00 & 1.00 & 1.00 & 1.00 & 1.00 & 1.00 & 1.00 & 1.00 & 0.99 & 1.00 \\
                \midrule[1.125pt]
                \bottomrule[1.5pt]
            \end{tabular}%
        }
    \end{center}
\end{table*}

\begin{table*}[!t]
    \begin{center}
        \caption{\label{tab:tnr-alpha-010} True Negative Rate (TNR) of all evaluated defense methods across datasets and malicious client ratios, shown for Dirichlet $\alpha=10$. Higher values indicate better preservation of benign updates.}
        \resizebox{\linewidth}{!}{%
            \begin{tabular}{ll g cgcg cgcg cgcg}\toprule[1.5pt]\midrule[1.125pt]
                & \multirow{2}{*}{\textbf{Baseline}} & \multicolumn{1}{c}{\multirow{2}{*}{\parbox{1cm}{\centering No Attack}}} & \multicolumn{4}{c}{$\epsilon = 10\%$} & \multicolumn{4}{c}{$\epsilon = 20\%$} & \multicolumn{4}{c}{$\epsilon = 30\%$} \\
                \cmidrule(lr){4-7} \cmidrule(lr){8-11} \cmidrule(lr){12-15}
                &  & \multicolumn{1}{c}{} & \multicolumn{1}{c}{\textbf{RN}} & \multicolumn{1}{c}{\textbf{LF}}  & \multicolumn{1}{c}{\textbf{SF}}  & \multicolumn{1}{c}{\textbf{IPM}} & \multicolumn{1}{c}{\textbf{RN}} & \multicolumn{1}{c}{\textbf{LF}}  & \multicolumn{1}{c}{\textbf{SF}}  & \multicolumn{1}{c}{\textbf{IPM}} & \multicolumn{1}{c}{\textbf{RN}} & \multicolumn{1}{c}{\textbf{LF}}  & \multicolumn{1}{c}{\textbf{SF}}  & \multicolumn{1}{c}{\textbf{IPM}} \\
				\midrule[1.125pt]
				\multirow{4}{*}{\rotatebox{90}{\textbf{MNIST}}}
				~ & \textsc{MultiKrum} & 1.00 & 0.96 & 0.96 & 0.94 & 0.95 & 0.90 & 0.91 & 0.91 & 0.85 & 0.71 & 0.83 & 0.89 & 0.70 \\
				~ & \textsc{GAN (Fixed Threshold)} & 0.99 & 0.99 & 0.99 & 0.98 & 1.00 & 0.98 & 1.00 & 0.99 & 1.00 & 0.99 & 0.98 & 1.00 & 0.98 \\
				~ & \textsc{GAN (Adaptive: Mean)} & 0.80 & 0.91 & 0.91 & 0.91 & 0.91 & 0.95 & 0.95 & 0.95 & 0.96 & 0.98 & 0.98 & 0.98 & 0.98 \\
				~ & \textsc{GAN (Clustered)} & 0.86 & 0.94 & 0.93 & 0.94 & 0.94 & 0.97 & 0.97 & 0.97 & 0.97 & 0.98 & 0.98 & 0.98 & 0.98 \\
				\midrule[1.125pt]
				\multirow{4}{*}{\rotatebox{90}{\textbf{FMNIST}}}
				~ & \textsc{MultiKrum} & 1.00 & 0.96 & 0.96 & 0.94 & 0.95 & 0.92 & 0.92 & 0.89 & 0.82 & 0.72 & 0.82 & 0.87 & 0.69 \\
				~ & \textsc{GAN (Fixed Threshold)} & 0.76 & 0.81 & 0.81 & 0.79 & 0.60 & 0.63 & 0.79 & 0.77 & 0.72 & 0.67 & 0.79 & 0.77 & 0.79 \\
				~ & \textsc{GAN (Adaptive: Mean)} & 0.72 & 0.79 & 0.78 & 0.76 & 0.76 & 0.81 & 0.80 & 0.80 & 0.82 & 0.82 & 0.81 & 0.81 & 0.84 \\
				~ & \textsc{GAN (Clustered)} & 0.78 & 0.82 & 0.83 & 0.81 & 0.81 & 0.85 & 0.82 & 0.84 & 0.83 & 0.87 & 0.86 & 0.87 & 0.85 \\
				\midrule[1.125pt]
				\multirow{4}{*}{\rotatebox{90}{\textbf{CIFAR-10}}}
				~ & \textsc{MultiKrum} & 1.00 & 0.96 & 0.96 & 0.96 & 0.95 & 0.92 & 0.93 & 0.93 & 0.92 & 0.86 & 0.89 & 0.91 & 0.79 \\
				~ & \textsc{GAN (Fixed Threshold)} & 0.03 & 0.03 & 0.03 & 0.03 & 0.03 & 0.03 & 0.03 & 0.03 & 0.03 & 0.03 & 0.03 & 0.03 & 0.03 \\
				~ & \textsc{GAN (Adaptive: Mean)} & 0.59 & 0.83 & 0.83 & 0.83 & 0.83 & 0.92 & 0.91 & 0.92 & 0.92 & 0.96 & 0.96 & 0.96 & 0.96 \\
				~ & \textsc{GAN (Clustered)} & 0.68 & 0.86 & 0.85 & 0.85 & 0.86 & 0.93 & 0.93 & 0.93 & 0.93 & 0.96 & 0.96 & 0.97 & 0.96 \\
                \midrule[1.125pt]
                \bottomrule[1.5pt]
            \end{tabular}%
        }
    \end{center}
\end{table*} % TPR/TNR for alpha = 10.0
\begin{table*}[!t]
    \begin{center}
        \caption{\label{tab:tpr-alpha-001} True Positive Rate (TPR) of all evaluated defense methods across datasets and malicious client ratios, shown for Dirichlet $\alpha=1.0$. Higher values indicate stronger detection of malicious updates.}
        \resizebox{\linewidth}{!}{%
            \begin{tabular}{ll g cgcg cgcg cgcg}\toprule[1.5pt]\midrule[1.125pt]
                & \multirow{2}{*}{\textbf{Baseline}} & \multicolumn{1}{c}{\multirow{2}{*}{\parbox{1cm}{\centering No Attack}}} & \multicolumn{4}{c}{$\epsilon = 10\%$} & \multicolumn{4}{c}{$\epsilon = 20\%$} & \multicolumn{4}{c}{$\epsilon = 30\%$} \\
                \cmidrule(lr){4-7} \cmidrule(lr){8-11} \cmidrule(lr){12-15}
                &  & \multicolumn{1}{c}{} & \multicolumn{1}{c}{\textbf{RN}} & \multicolumn{1}{c}{\textbf{LF}}  & \multicolumn{1}{c}{\textbf{SF}}  & \multicolumn{1}{c}{\textbf{IPM}} & \multicolumn{1}{c}{\textbf{RN}} & \multicolumn{1}{c}{\textbf{LF}}  & \multicolumn{1}{c}{\textbf{SF}}  & \multicolumn{1}{c}{\textbf{IPM}} & \multicolumn{1}{c}{\textbf{RN}} & \multicolumn{1}{c}{\textbf{LF}}  & \multicolumn{1}{c}{\textbf{SF}}  & \multicolumn{1}{c}{\textbf{IPM}} \\
				\midrule[1.125pt]
				\multirow{4}{*}{\rotatebox{90}{\textbf{MNIST}}}
				~ & \textsc{MultiKrum} & 0.00 & 0.68 & 0.67 & 0.67 & 0.62 & 0.60 & 0.74 & 0.68 & 0.33 & 0.34 & 0.60 & 0.72 & 0.29 \\
				~ & \textsc{GAN (Fixed Threshold)} & 0.00 & 1.00 & 1.00 & 1.00 & 1.00 & 1.00 & 1.00 & 1.00 & 1.00 & 1.00 & 1.00 & 1.00 & 1.00 \\
				~ & \textsc{GAN (Adaptive: Mean)} & 0.00 & 1.00 & 1.00 & 1.00 & 1.00 & 1.00 & 1.00 & 1.00 & 1.00 & 1.00 & 1.00 & 1.00 & 1.00 \\
				~ & \textsc{GAN (Clustered)} & 0.00 & 1.00 & 1.00 & 1.00 & 1.00 & 1.00 & 1.00 & 1.00 & 1.00 & 1.00 & 1.00 & 1.00 & 1.00 \\
				\midrule[1.125pt]
				\multirow{4}{*}{\rotatebox{90}{\textbf{FMNIST}}}
				~ & \textsc{MultiKrum} & 0.00 & 0.68 & 0.67 & 0.63 & 0.60 & 0.71 & 0.74 & 0.56 & 0.32 & 0.32 & 0.66 & 0.64 & 0.30 \\
				~ & \textsc{GAN (Fixed Threshold)} & 0.00 & 1.00 & 1.00 & 1.00 & 1.00 & 1.00 & 1.00 & 1.00 & 1.00 & 1.00 & 1.00 & 1.00 & 1.00 \\
				~ & \textsc{GAN (Adaptive: Mean)} & 0.00 & 1.00 & 1.00 & 1.00 & 1.00 & 1.00 & 1.00 & 1.00 & 1.00 & 1.00 & 1.00 & 1.00 & 1.00 \\
				~ & \textsc{GAN (Clustered)} & 0.00 & 1.00 & 1.00 & 1.00 & 1.00 & 1.00 & 0.95 & 1.00 & 1.00 & 1.00 & 0.94 & 1.00 & 1.00 \\
				\midrule[1.125pt]
				\multirow{4}{*}{\rotatebox{90}{\textbf{CIFAR-10}}}
				~ & \textsc{MultiKrum} & 0.00 & 0.68 & 0.68 & 0.68 & 0.65 & 0.73 & 0.75 & 0.76 & 0.72 & 0.70 & 0.82 & 0.83 & 0.49 \\
				~ & \textsc{GAN (Fixed Threshold)} & 0.00 & 1.00 & 1.00 & 1.00 & 1.00 & 1.00 & 1.00 & 1.00 & 1.00 & 1.00 & 1.00 & 1.00 & 1.00 \\
				~ & \textsc{GAN (Adaptive: Mean)} & 0.00 & 1.00 & 1.00 & 1.00 & 1.00 & 1.00 & 0.99 & 1.00 & 1.00 & 1.00 & 0.98 & 0.98 & 1.00 \\
				~ & \textsc{GAN (Clustered)} & 0.00 & 1.00 & 0.96 & 0.97 & 1.00 & 1.00 & 0.96 & 0.94 & 1.00 & 1.00 & 0.94 & 0.86 & 1.00 \\
                \midrule[1.125pt]
                \bottomrule[1.5pt]
            \end{tabular}%
        }
    \end{center}
\end{table*}

\begin{table*}[!t]
    \begin{center}
        \caption{\label{tab:tnr-alpha-001} True Negative Rate (TNR) of all evaluated defense methods across datasets and malicious client ratios, shown for Dirichlet $\alpha=1.0$. Higher values indicate better preservation of benign updates.}
        \resizebox{\linewidth}{!}{%
            \begin{tabular}{ll g cgcg cgcg cgcg}\toprule[1.5pt]\midrule[1.125pt]
                & \multirow{2}{*}{\textbf{Baseline}} & \multicolumn{1}{c}{\multirow{2}{*}{\parbox{1cm}{\centering No Attack}}} & \multicolumn{4}{c}{$\epsilon = 10\%$} & \multicolumn{4}{c}{$\epsilon = 20\%$} & \multicolumn{4}{c}{$\epsilon = 30\%$} \\
                \cmidrule(lr){4-7} \cmidrule(lr){8-11} \cmidrule(lr){12-15}
                &  & \multicolumn{1}{c}{} & \multicolumn{1}{c}{\textbf{RN}} & \multicolumn{1}{c}{\textbf{LF}}  & \multicolumn{1}{c}{\textbf{SF}}  & \multicolumn{1}{c}{\textbf{IPM}} & \multicolumn{1}{c}{\textbf{RN}} & \multicolumn{1}{c}{\textbf{LF}}  & \multicolumn{1}{c}{\textbf{SF}}  & \multicolumn{1}{c}{\textbf{IPM}} & \multicolumn{1}{c}{\textbf{RN}} & \multicolumn{1}{c}{\textbf{LF}}  & \multicolumn{1}{c}{\textbf{SF}}  & \multicolumn{1}{c}{\textbf{IPM}} \\
				\midrule[1.125pt]
				\multirow{4}{*}{\rotatebox{90}{\textbf{MNIST}}}
				~ & \textsc{MultiKrum} & 1.00 & 0.96 & 0.96 & 0.96 & 0.95 & 0.89 & 0.92 & 0.91 & 0.83 & 0.71 & 0.82 & 0.86 & 0.70 \\
				~ & \textsc{GAN (Fixed Threshold)} & 0.98 & 0.98 & 0.98 & 0.98 & 0.98 & 0.98 & 0.98 & 0.98 & 0.98 & 0.98 & 0.98 & 0.97 & 0.98 \\
				~ & \textsc{GAN (Adaptive: Mean)} & 0.83 & 0.91 & 0.91 & 0.91 & 0.92 & 0.95 & 0.95 & 0.95 & 0.95 & 0.97 & 0.97 & 0.97 & 0.97 \\
				~ & \textsc{GAN (Clustered)} & 0.86 & 0.93 & 0.94 & 0.93 & 0.94 & 0.96 & 0.96 & 0.96 & 0.96 & 0.98 & 0.97 & 0.97 & 0.98 \\
				\midrule[1.125pt]
				\multirow{4}{*}{\rotatebox{90}{\textbf{FMNIST}}}
				~ & \textsc{MultiKrum} & 1.00 & 0.96 & 0.96 & 0.95 & 0.95 & 0.92 & 0.93 & 0.88 & 0.83 & 0.71 & 0.84 & 0.84 & 0.70 \\
				~ & \textsc{GAN (Fixed Threshold)} & 0.11 & 0.15 & 0.27 & 0.15 & 0.11 & 0.31 & 0.44 & 0.36 & 0.23 & 0.19 & 0.13 & 0.33 & 0.27 \\
				~ & \textsc{GAN (Adaptive: Mean)} & 0.62 & 0.64 & 0.64 & 0.66 & 0.64 & 0.70 & 0.69 & 0.69 & 0.68 & 0.73 & 0.74 & 0.72 & 0.73 \\
				~ & \textsc{GAN (Clustered)} & 0.65 & 0.67 & 0.69 & 0.70 & 0.66 & 0.69 & 0.78 & 0.71 & 0.68 & 0.70 & 0.73 & 0.70 & 0.72 \\
				\midrule[1.125pt]
				\multirow{4}{*}{\rotatebox{90}{\textbf{CIFAR-10}}}
				~ & \textsc{MultiKrum} & 1.00 & 0.96 & 0.96 & 0.96 & 0.96 & 0.92 & 0.93 & 0.93 & 0.92 & 0.86 & 0.90 & 0.91 & 0.78 \\
				~ & \textsc{GAN (Fixed Threshold)} & 0.03 & 0.03 & 0.03 & 0.03 & 0.03 & 0.03 & 0.03 & 0.03 & 0.03 & 0.03 & 0.03 & 0.03 & 0.03 \\
				~ & \textsc{GAN (Adaptive: Mean)} & 0.57 & 0.83 & 0.76 & 0.81 & 0.83 & 0.92 & 0.86 & 0.90 & 0.92 & 0.96 & 0.92 & 0.95 & 0.96 \\
				~ & \textsc{GAN (Clustered)} & 0.66 & 0.86 & 0.85 & 0.86 & 0.87 & 0.94 & 0.93 & 0.94 & 0.93 & 0.97 & 0.96 & 0.97 & 0.96 \\
                \midrule[1.125pt]
                \bottomrule[1.5pt]
            \end{tabular}%
        }
    \end{center}
\end{table*} % TPR/TNR for alpha =  1.0
\begin{table*}[!t]
    \begin{center}
        \caption{\label{tab:tpr-alpha-000_1} True Positive Rate (TPR) of all evaluated defense methods across datasets and malicious client ratios, shown for Dirichlet $\alpha=0.1$. Higher values indicate stronger detection of malicious updates.}
        \resizebox{\linewidth}{!}{%
            \begin{tabular}{ll g cgcg cgcg cgcg}\toprule[1.5pt]\midrule[1.125pt]
                & \multirow{2}{*}{\textbf{Baseline}} & \multicolumn{1}{c}{\multirow{2}{*}{\parbox{1cm}{\centering No Attack}}} & \multicolumn{4}{c}{$\epsilon = 10\%$} & \multicolumn{4}{c}{$\epsilon = 20\%$} & \multicolumn{4}{c}{$\epsilon = 30\%$} \\
                \cmidrule(lr){4-7} \cmidrule(lr){8-11} \cmidrule(lr){12-15}
                &  & \multicolumn{1}{c}{} & \multicolumn{1}{c}{\textbf{RN}} & \multicolumn{1}{c}{\textbf{LF}}  & \multicolumn{1}{c}{\textbf{SF}}  & \multicolumn{1}{c}{\textbf{IPM}} & \multicolumn{1}{c}{\textbf{RN}} & \multicolumn{1}{c}{\textbf{LF}}  & \multicolumn{1}{c}{\textbf{SF}}  & \multicolumn{1}{c}{\textbf{IPM}} & \multicolumn{1}{c}{\textbf{RN}} & \multicolumn{1}{c}{\textbf{LF}}  & \multicolumn{1}{c}{\textbf{SF}}  & \multicolumn{1}{c}{\textbf{IPM}} \\
				\midrule[1.125pt]
				\multirow{4}{*}{\rotatebox{90}{\textbf{MNIST}}}
				~ & \textsc{MultiKrum} & 0.00 & 0.68 & 0.59 & 0.49 & 0.65 & 0.63 & 0.66 & 0.71 & 0.31 & 0.31 & 0.68 & 0.72 & 0.29 \\
				~ & \textsc{GAN (Fixed Threshold)} & 0.00 & 1.00 & 1.00 & 1.00 & 1.00 & 1.00 & 1.00 & 1.00 & 1.00 & 1.00 & 1.00 & 1.00 & 1.00 \\
				~ & \textsc{GAN (Adaptive: Mean)} & 0.00 & 0.98 & 0.96 & 0.99 & 1.00 & 0.99 & 0.92 & 0.99 & 1.00 & 0.99 & 0.88 & 0.98 & 1.00 \\
				~ & \textsc{GAN (Clustered)} & 0.00 & 0.98 & 0.89 & 0.98 & 1.00 & 0.98 & 0.87 & 0.98 & 1.00 & 0.99 & 0.82 & 0.97 & 1.00 \\
				\midrule[1.125pt]
				\multirow{4}{*}{\rotatebox{90}{\textbf{FMNIST}}}
				~ & \textsc{MultiKrum} & 0.00 & 0.68 & 0.61 & 0.56 & 0.53 & 0.59 & 0.70 & 0.33 & 0.20 & 0.32 & 0.71 & 0.82 & 0.29 \\
				~ & \textsc{GAN (Fixed Threshold)} & 0.00 & 1.00 & 1.00 & 1.00 & 1.00 & 1.00 & 1.00 & 1.00 & 1.00 & 1.00 & 1.00 & 1.00 & 1.00 \\
				~ & \textsc{GAN (Adaptive: Mean)} & 0.00 & 0.99 & 0.96 & 1.00 & 1.00 & 0.99 & 0.96 & 1.00 & 1.00 & 0.99 & 0.95 & 1.00 & 1.00 \\
				~ & \textsc{GAN (Clustered)} & 0.00 & 0.97 & 0.98 & 0.96 & 1.00 & 0.98 & 0.97 & 0.96 & 1.00 & 0.99 & 0.97 & 0.97 & 1.00 \\
				\midrule[1.125pt]
				\multirow{4}{*}{\rotatebox{90}{\textbf{CIFAR-10}}}
				~ & \textsc{MultiKrum} & 0.00 & 0.68 & 0.66 & 0.68 & 0.67 & 0.73 & 0.74 & 0.76 & 0.73 & 0.70 & 0.80 & 0.83 & 0.53 \\
				~ & \textsc{GAN (Fixed Threshold)} & 0.00 & 1.00 & 1.00 & 1.00 & 1.00 & 1.00 & 1.00 & 1.00 & 1.00 & 1.00 & 1.00 & 1.00 & 1.00 \\
				~ & \textsc{GAN (Adaptive: Mean)} & 0.00 & 1.00 & 0.94 & 0.97 & 1.00 & 1.00 & 0.91 & 0.94 & 1.00 & 1.00 & 0.84 & 0.88 & 1.00 \\
				~ & \textsc{GAN (Clustered)} & 0.00 & 1.00 & 0.87 & 0.90 & 1.00 & 1.00 & 0.77 & 0.81 & 1.00 & 1.00 & 0.69 & 0.74 & 1.00 \\
                \midrule[1.125pt]
                \bottomrule[1.5pt]
            \end{tabular}%
        }
    \end{center}
\end{table*}

\begin{table*}[!t]
    \begin{center}
        \caption{\label{tab:tnr-alpha-000_1} True Negative Rate (TNR) of all evaluated defense methods across datasets and malicious client ratios, shown for Dirichlet $\alpha=0.1$. Higher values indicate better preservation of benign updates.}
        \resizebox{\linewidth}{!}{%
            \begin{tabular}{ll g cgcg cgcg cgcg}\toprule[1.5pt]\midrule[1.125pt]
                & \multirow{2}{*}{\textbf{Baseline}} & \multicolumn{1}{c}{\multirow{2}{*}{\parbox{1cm}{\centering No Attack}}} & \multicolumn{4}{c}{$\epsilon = 10\%$} & \multicolumn{4}{c}{$\epsilon = 20\%$} & \multicolumn{4}{c}{$\epsilon = 30\%$} \\
                \cmidrule(lr){4-7} \cmidrule(lr){8-11} \cmidrule(lr){12-15}
                &  & \multicolumn{1}{c}{} & \multicolumn{1}{c}{\textbf{RN}} & \multicolumn{1}{c}{\textbf{LF}}  & \multicolumn{1}{c}{\textbf{SF}}  & \multicolumn{1}{c}{\textbf{IPM}} & \multicolumn{1}{c}{\textbf{RN}} & \multicolumn{1}{c}{\textbf{LF}}  & \multicolumn{1}{c}{\textbf{SF}}  & \multicolumn{1}{c}{\textbf{IPM}} & \multicolumn{1}{c}{\textbf{RN}} & \multicolumn{1}{c}{\textbf{LF}}  & \multicolumn{1}{c}{\textbf{SF}}  & \multicolumn{1}{c}{\textbf{IPM}} \\
				\midrule[1.125pt]
				\multirow{4}{*}{\rotatebox{90}{\textbf{MNIST}}}
				~ & \textsc{MultiKrum} & 1.00 & 0.96 & 0.95 & 0.94 & 0.96 & 0.90 & 0.91 & 0.92 & 0.83 & 0.71 & 0.85 & 0.86 & 0.69 \\
				~ & \textsc{GAN (Fixed Threshold)} & 0.03 & 0.03 & 0.03 & 0.03 & 0.03 & 0.03 & 0.03 & 0.03 & 0.03 & 0.03 & 0.03 & 0.03 & 0.03 \\
				~ & \textsc{GAN (Adaptive: Mean)} & 0.61 & 0.73 & 0.72 & 0.73 & 0.79 & 0.76 & 0.76 & 0.82 & 0.82 & 0.80 & 0.81 & 0.88 & 0.91 \\
				~ & \textsc{GAN (Clustered)} & 0.73 & 0.78 & 0.81 & 0.82 & 0.84 & 0.70 & 0.81 & 0.89 & 0.90 & 0.89 & 0.86 & 0.94 & 0.95 \\
				\midrule[1.125pt]
				\multirow{4}{*}{\rotatebox{90}{\textbf{FMNIST}}}
				~ & \textsc{MultiKrum} & 1.00 & 0.96 & 0.95 & 0.95 & 0.94 & 0.89 & 0.91 & 0.83 & 0.80 & 0.71 & 0.86 & 0.91 & 0.70 \\
				~ & \textsc{GAN (Fixed Threshold)} & 0.03 & 0.02 & 0.03 & 0.03 & 0.03 & 0.03 & 0.03 & 0.03 & 0.03 & 0.03 & 0.03 & 0.03 & 0.02 \\
				~ & \textsc{GAN (Adaptive: Mean)} & 0.36 & 0.39 & 0.38 & 0.39 & 0.39 & 0.39 & 0.41 & 0.43 & 0.42 & 0.37 & 0.45 & 0.38 & 0.20 \\
				~ & \textsc{GAN (Clustered)} & 0.26 & 0.63 & 0.28 & 0.39 & 0.28 & 0.68 & 0.29 & 0.55 & 0.56 & 0.73 & 0.32 & 0.58 & 0.97 \\
				\midrule[1.125pt]
				\multirow{4}{*}{\rotatebox{90}{\textbf{CIFAR-10}}}
				~ & \textsc{MultiKrum} & 1.00 & 0.96 & 0.96 & 0.96 & 0.96 & 0.92 & 0.92 & 0.93 & 0.92 & 0.86 & 0.90 & 0.91 & 0.79 \\
				~ & \textsc{GAN (Fixed Threshold)} & 0.03 & 0.03 & 0.03 & 0.03 & 0.03 & 0.03 & 0.03 & 0.03 & 0.03 & 0.03 & 0.03 & 0.03 & 0.03 \\
				~ & \textsc{GAN (Adaptive: Mean)} & 0.57 & 0.76 & 0.68 & 0.76 & 0.82 & 0.86 & 0.75 & 0.86 & 0.91 & 0.92 & 0.83 & 0.93 & 0.96 \\
				~ & \textsc{GAN (Clustered)} & 0.64 & 0.84 & 0.78 & 0.84 & 0.84 & 0.92 & 0.84 & 0.92 & 0.93 & 0.96 & 0.91 & 0.96 & 0.97 \\
                \midrule[1.125pt]
                \bottomrule[1.5pt]
            \end{tabular}%
        }
    \end{center}
\end{table*} % TPR/TNR for alpha =  0.1

\section{Additional Results}
\label{sec:appendix-b}
In this appendix, we provide comprehensive detection results of our framework across all datasets, malicious ratios, attack types, and data heterogeneity levels (Dirichlet $\alpha$ values). To avoid overwhelming the main body, only one representative set of results is presented there (e.g., CIFAR-10, $\alpha=1.0$), while the full suite is summarized here. 

Tables~\ref{tab:tpr-alpha-100}--\ref{tab:tnr-alpha-000_1} report the True Positive Rate (TPR) and True Negative Rate (TNR) for all combinations of datasets (MNIST, Fashion-MNIST, CIFAR-10), fractions of malicious clients ($0\%$, $10\%$, $20\%$, $30\%$), and evaluated defense methods under four attack types (Random Noise, Sign-Flipping, Label-Flipping, Inner Product Manipulation). Each Dirichlet $\alpha \in \{100.0, 10.0, 1.0, 0.1\}$  corresponds to a separate table pair (TPR and TNR), capturing both IID and non-IID regimes.

These tables highlight several key patterns:
\begin{itemize}
    \item Our adaptive-threshold and clustering-based GAN variants consistently achieve high TPR across datasets and attack types, demonstrating strong detection of malicious updates.
    \item Benign contributions are preserved effectively, reflected in consistently high TNR values across all scenarios.
    \item The impact of data heterogeneity is visible: as $\alpha$ decreases (more non-IID), TPR/TNR can slightly vary, emphasizing the importance of adaptive defenses.
    \item Comparisons with baseline aggregation schemes such as \textsc{MultiKrum}, \textsc{Median}, and \textsc{TrimAvg} reveal substantial improvements in adversarial resilience, especially under high malicious client ratios.
\end{itemize}

\begin{table*}[!ht]
    % {\centering
    \begin{center}
        \caption{\label{tab:results-alpha-010} Overall test accuracy (ACC) of federated learning under benign training and poisoning attacks across malicious client fractions (\(10\%\), \(20\%\), \(30\%\)) in the low non-IID setting \(\alpha = 10\). Results are reported on MNIST, Fashion-MNIST, and CIFAR-10, comparing standard aggregation rules (\textsc{FedAvg, Median, Trimmed Mean, GeoMedian, Krum, NNM+Krum}) against our proposed GAN-based authentication variants. GAN-based defenses consistently preserve accuracy even under high adversarial presence, whereas conventional robust aggregation methods degrade significantly.}
        
        \resizebox{1.0\linewidth}{!}{%
            \begin{tabular}{ll g cgcg cgcg cgcg}\toprule[1.5pt]\midrule[1.125pt]
                & \multirow{2}{*}{\textbf{Baseline}} & \multicolumn{1}{c}{\multirow{2}{*}{\parbox{1cm}{\centering No Attack}}} & \multicolumn{4}{c}{$\epsilon = 10\%$} & \multicolumn{4}{c}{$\epsilon = 20\%$} & \multicolumn{4}{c}{$\epsilon = 30\%$} \\
                \cmidrule(lr){4-7} \cmidrule(lr){8-11} \cmidrule(lr){12-15}
                &  & \multicolumn{1}{c}{} & \multicolumn{1}{c}{\textbf{RN}} & \multicolumn{1}{c}{\textbf{LF}}  & \multicolumn{1}{c}{\textbf{SF}}  & \multicolumn{1}{c}{\textbf{IPM}} & \multicolumn{1}{c}{\textbf{RN}} & \multicolumn{1}{c}{\textbf{LF}}  & \multicolumn{1}{c}{\textbf{SF}}  & \multicolumn{1}{c}{\textbf{IPM}} & \multicolumn{1}{c}{\textbf{RN}} & \multicolumn{1}{c}{\textbf{LF}}  & \multicolumn{1}{c}{\textbf{SF}}  & \multicolumn{1}{c}{\textbf{IPM}} \\
				\midrule[1.125pt]
				\multirow{8}{*}{\rotatebox{90}{\textbf{MNIST}}}
				  & \textsc{FedAvg} & 0.99 & 0.47 & 0.98 & 0.11 & 0.11 & 0.52 & 0.11 & 0.10 & 0.11 & 0.13 & 0.10 & 0.10 & 0.10 \\
				~ & \textsc{Median} & 0.99 & 0.99 & 0.99 & 0.99 & 0.99 & 0.94 & 0.11 & 0.99 & 0.10 & 0.10 & 0.11 & 0.11 & 0.10 \\
				~ & \textsc{TrimAvg} & 0.99 & 0.93 & 0.99 & 0.98 & 0.10 & 0.40 & 0.98 & 0.11 & 0.10 & 0.10 & 0.11 & 0.11 & 0.10 \\
				~ & \textsc{GeoMedian} & 0.99 & 0.99 & 0.99 & 0.99 & 0.99 & 0.95 & 0.99 & 0.99 & 0.10 & 0.10 & 0.10 & 0.11 & 0.10 \\
				~ & \textsc{MultiKrum} & 0.99 & 0.93 & 0.99 & 0.98 & 0.11 & 0.81 & 0.98 & 0.97 & 0.10 & 0.10 & 0.11 & 0.81 & 0.10 \\
				~ & \textsc{NNM+Krum} & 0.99 & 0.92 & 0.99 & 0.98 & 0.10 & 0.89 & 0.11 & 0.11 & 0.10 & 0.10 & 0.11 & 0.11 & 0.10 \\ \cmidrule(lr){2-15}
				~ & \textsc{GAN (Fixed Threshold)} & 0.99 & 0.99 & 0.99 & 0.99 & 0.99 & 0.99 & 0.99 & 0.99 & 0.99 & 0.99 & 0.99 & 0.99 & 0.99 \\
				~ & \textsc{GAN (Adaptive: Mean)} & 0.99 & 0.99 & 0.99 & 0.99 & 0.99 & 0.99 & 0.99 & 0.99 & 0.99 & 0.99 & 0.99 & 0.99 & 0.99 \\
				~ & \textsc{GAN (Clustered)} & 0.99 & 0.99 & 0.99 & 0.99 & 0.99 & 0.99 & 0.99 & 0.99 & 0.99 & 0.99 & 0.99 & 0.99 & 0.99 \\
				\midrule[1.125pt]
				\multirow{8}{*}{\rotatebox{90}{\textbf{Fashion MNIST}}}
				  & \textsc{FedAvg} & 0.89 & 0.19 & 0.88 & 0.10 & 0.10 & 0.30 & 0.10 & 0.10 & 0.10 & 0.15 & 0.35 & 0.10 & 0.10 \\
				~ & \textsc{Median} & 0.89 & 0.88 & 0.88 & 0.88 & 0.88 & 0.68 & 0.78 & 0.86 & 0.10 & 0.10 & 0.10 & 0.10 & 0.10 \\
				~ & \textsc{TrimAvg} & 0.89 & 0.69 & 0.88 & 0.78 & 0.10 & 0.68 & 0.82 & 0.10 & 0.10 & 0.10 & 0.10 & 0.10 & 0.10 \\
				~ & \textsc{GeoMedian} & 0.89 & 0.88 & 0.89 & 0.88 & 0.89 & 0.81 & 0.85 & 0.87 & 0.10 & 0.10 & 0.10 & 0.10 & 0.10 \\
				~ & \textsc{MultiKrum} & 0.89 & 0.78 & 0.89 & 0.84 & 0.10 & 0.12 & 0.87 & 0.10 & 0.10 & 0.10 & 0.10 & 0.79 & 0.10 \\
				~ & \textsc{NNM+Krum} & 0.89 & 0.74 & 0.89 & 0.10 & 0.10 & 0.70 & 0.84 & 0.10 & 0.10 & 0.10 & 0.10 & 0.10 & 0.10 \\ \cmidrule(lr){2-15}
				~ & \textsc{GAN (Fixed Threshold)} & 0.89 & 0.89 & 0.89 & 0.89 & 0.89 & 0.89 & 0.89 & 0.89 & 0.89 & 0.89 & 0.89 & 0.89 & 0.89 \\
				~ & \textsc{GAN (Adaptive: Mean)} & 0.90 & 0.89 & 0.89 & 0.89 & 0.89 & 0.89 & 0.89 & 0.89 & 0.89 & 0.89 & 0.89 & 0.88 & 0.89 \\
				~ & \textsc{GAN (Clustered)} & 0.89 & 0.89 & 0.89 & 0.89 & 0.89 & 0.89 & 0.89 & 0.89 & 0.89 & 0.89 & 0.89 & 0.88 & 0.89 \\
				\midrule[1.125pt]
				\multirow{8}{*}{\rotatebox{90}{\textbf{CIFAR-10}}}
				  & \textsc{FedAvg} & 0.88 & 0.12 & 0.87 & 0.57 & 0.10 & 0.11 & 0.86 & 0.11 & 0.10 & 0.11 & 0.62 & 0.10 & 0.10 \\
				~ & \textsc{Median} & 0.88 & 0.87 & 0.87 & 0.86 & 0.86 & 0.42 & 0.86 & 0.81 & 0.10 & 0.18 & 0.84 & 0.72 & 0.10 \\
				~ & \textsc{TrimAvg} & 0.89 & 0.13 & 0.88 & 0.81 & 0.23 & 0.12 & 0.86 & 0.76 & 0.66 & 0.12 & 0.84 & 0.61 & 0.10 \\
				~ & \textsc{GeoMedian} & 0.89 & 0.88 & 0.88 & 0.86 & 0.86 & 0.44 & 0.87 & 0.82 & 0.19 & 0.20 & 0.86 & 0.76 & 0.10 \\
				~ & \textsc{MultiKrum} & 0.88 & 0.13 & 0.88 & 0.82 & 0.10 & 0.12 & 0.87 & 0.80 & 0.10 & 0.12 & 0.85 & 0.73 & 0.33 \\
				~ & \textsc{NNM+Krum} & 0.89 & 0.13 & 0.88 & 0.83 & 0.10 & 0.12 & 0.87 & 0.79 & 0.29 & 0.12 & 0.84 & 0.72 & 0.10 \\ \cmidrule(lr){2-15}
				~ & \textsc{GAN (Fixed Threshold)} & 0.56 & 0.56 & 0.57 & 0.56 & 0.57 & 0.56 & 0.57 & 0.57 & 0.56 & 0.56 & 0.56 & 0.55 & 0.56 \\
				~ & \textsc{GAN (Adaptive: Mean)} & 0.87 & 0.88 & 0.88 & 0.88 & 0.88 & 0.88 & 0.87 & 0.88 & 0.88 & 0.87 & 0.87 & 0.87 & 0.87 \\
				~ & \textsc{GAN (Clustered)} & 0.88 & 0.88 & 0.88 & 0.88 & 0.88 & 0.88 & 0.87 & 0.87 & 0.88 & 0.87 & 0.87 & 0.87 & 0.87 \\
                \midrule[1.125pt]
                \bottomrule[1.5pt]
            \end{tabular}%
        }
    \end{center}
    \footnotesize{*we set $\beta =$ $0.1$, $0.2$, $0.3$ (where $\beta$ is byzantine aggregation heuristic, for \textsc{TrimAvg} and \textsc{MultiKrum} respectively under $10\%$, $20\%$, and $30\%$ malicious clients.}
\end{table*}  % ACC for alpha = 10.0
\begin{table*}[!ht]
    % {\centering
    \begin{center}
        \caption{\label{tab:results-alpha-001} Overall test accuracy (ACC) of federated learning under benign training and poisoning attacks across malicious client fractions (\(10\%\), \(20\%\), \(30\%\)) in the moderate non-IID setting \(\alpha = 1.0\). Results are reported on MNIST, Fashion-MNIST, and CIFAR-10, comparing standard aggregation rules (\textsc{FedAvg, Median, Trimmed Mean, GeoMedian, Krum, NNM+Krum}) against our proposed GAN-based authentication variants. GAN-based defenses consistently preserve accuracy even under high adversarial presence, whereas conventional robust aggregation methods degrade significantly.}

        \resizebox{1.0\linewidth}{!}{%
            \begin{tabular}{ll g cgcg cgcg cgcg}\toprule[1.5pt]\midrule[1.125pt]
                & \multirow{2}{*}{\textbf{Baseline}} & \multicolumn{1}{c}{\multirow{2}{*}{\parbox{1cm}{\centering No Attack}}} & \multicolumn{4}{c}{$\epsilon = 10\%$} & \multicolumn{4}{c}{$\epsilon = 20\%$} & \multicolumn{4}{c}{$\epsilon = 30\%$} \\
                \cmidrule(lr){4-7} \cmidrule(lr){8-11} \cmidrule(lr){12-15}
                &  & \multicolumn{1}{c}{} & \multicolumn{1}{c}{\textbf{RN}} & \multicolumn{1}{c}{\textbf{LF}}  & \multicolumn{1}{c}{\textbf{SF}}  & \multicolumn{1}{c}{\textbf{IPM}} & \multicolumn{1}{c}{\textbf{RN}} & \multicolumn{1}{c}{\textbf{LF}}  & \multicolumn{1}{c}{\textbf{SF}}  & \multicolumn{1}{c}{\textbf{IPM}} & \multicolumn{1}{c}{\textbf{RN}} & \multicolumn{1}{c}{\textbf{LF}}  & \multicolumn{1}{c}{\textbf{SF}}  & \multicolumn{1}{c}{\textbf{IPM}} \\
				\midrule[1.125pt]
				\multirow{8}{*}{\rotatebox{90}{\textbf{MNIST}}}
				  & \textsc{FedAvg} & 0.99 & 0.88 & 0.99 & 0.11 & 0.10 & 0.62 & 0.96 & 0.10 & 0.11 & 0.10 & 0.11 & 0.10 & 0.10 \\
				~ & \textsc{Median} & 0.99 & 0.99 & 0.99 & 0.99 & 0.99 & 0.92 & 0.98 & 0.98 & 0.10 & 0.10 & 0.10 & 0.11 & 0.10 \\
				~ & \textsc{TrimAvg} & 0.99 & 0.87 & 0.99 & 0.98 & 0.10 & 0.81 & 0.98 & 0.10 & 0.10 & 0.10 & 0.11 & 0.11 & 0.10 \\
				~ & \textsc{GeoMedian} & 0.99 & 0.99 & 0.99 & 0.99 & 0.99 & 0.94 & 0.99 & 0.99 & 0.10 & 0.10 & 0.11 & 0.11 & 0.10 \\
				~ & \textsc{MultiKrum} & 0.99 & 0.92 & 0.99 & 0.97 & 0.10 & 0.78 & 0.99 & 0.93 & 0.10 & 0.10 & 0.11 & 0.11 & 0.10 \\
				~ & \textsc{NNM+Krum} & 0.99 & 0.92 & 0.99 & 0.94 & 0.10 & 0.85 & 0.98 & 0.10 & 0.10 & 0.10 & 0.11 & 0.11 & 0.10 \\ \cmidrule(lr){2-15}
				~ & \textsc{GAN (Fixed Threshold)} & 0.99 & 0.99 & 0.99 & 0.99 & 0.99 & 0.99 & 0.99 & 0.99 & 0.99 & 0.99 & 0.99 & 0.99 & 0.99 \\
				~ & \textsc{GAN (Adaptive: Mean)} & 0.99 & 0.99 & 0.99 & 0.99 & 0.99 & 0.99 & 0.99 & 0.99 & 0.99 & 0.99 & 0.99 & 0.99 & 0.99 \\
				~ & \textsc{GAN (Clustered)} & 0.99 & 0.99 & 0.99 & 0.99 & 0.99 & 0.99 & 0.99 & 0.99 & 0.99 & 0.99 & 0.99 & 0.99 & 0.99 \\
				\midrule[1.125pt]
				\multirow{8}{*}{\rotatebox{90}{\textbf{Fashion MNIST}}}
				  & \textsc{FedAvg} & 0.88 & 0.71 & 0.86 & 0.10 & 0.10 & 0.21 & 0.81 & 0.10 & 0.10 & 0.10 & 0.27 & 0.10 & 0.10 \\
				~ & \textsc{Median} & 0.88 & 0.88 & 0.87 & 0.87 & 0.88 & 0.67 & 0.85 & 0.85 & 0.10 & 0.10 & 0.10 & 0.10 & 0.10 \\
				~ & \textsc{TrimAvg} & 0.89 & 0.60 & 0.87 & 0.85 & 0.10 & 0.67 & 0.85 & 0.10 & 0.10 & 0.10 & 0.80 & 0.10 & 0.10 \\
				~ & \textsc{GeoMedian} & 0.88 & 0.88 & 0.88 & 0.88 & 0.88 & 0.79 & 0.86 & 0.86 & 0.10 & 0.10 & 0.10 & 0.10 & 0.10 \\
				~ & \textsc{MultiKrum} & 0.89 & 0.74 & 0.88 & 0.10 & 0.10 & 0.64 & 0.87 & 0.77 & 0.10 & 0.10 & 0.10 & 0.10 & 0.10 \\
				~ & \textsc{NNM+Krum} & 0.89 & 0.50 & 0.88 & 0.75 & 0.10 & 0.70 & 0.87 & 0.10 & 0.10 & 0.10 & 0.10 & 0.10 & 0.10 \\ \cmidrule(lr){2-15}
				~ & \textsc{GAN (Fixed Threshold)} & 0.87 & 0.88 & 0.88 & 0.87 & 0.87 & 0.88 & 0.87 & 0.87 & 0.87 & 0.87 & 0.87 & 0.87 & 0.87 \\
				~ & \textsc{GAN (Adaptive: Mean)} & 0.88 & 0.88 & 0.88 & 0.88 & 0.88 & 0.88 & 0.88 & 0.88 & 0.88 & 0.88 & 0.87 & 0.88 & 0.88 \\
				~ & \textsc{GAN (Clustered)} & 0.89 & 0.88 & 0.88 & 0.88 & 0.88 & 0.88 & 0.88 & 0.88 & 0.88 & 0.88 & 0.88 & 0.88 & 0.88 \\
				\midrule[1.125pt]
				\multirow{8}{*}{\rotatebox{90}{\textbf{CIFAR-10}}}
				  & \textsc{FedAvg} & 0.87 & 0.11 & 0.86 & 0.34 & 0.10 & 0.11 & 0.84 & 0.12 & 0.10 & 0.11 & 0.62 & 0.10 & 0.10 \\
				~ & \textsc{Median} & 0.87 & 0.85 & 0.85 & 0.85 & 0.84 & 0.37 & 0.84 & 0.80 & 0.10 & 0.16 & 0.82 & 0.70 & 0.10 \\
				~ & \textsc{TrimAvg} & 0.87 & 0.12 & 0.86 & 0.80 & 0.22 & 0.12 & 0.85 & 0.71 & 0.51 & 0.14 & 0.81 & 0.55 & 0.10 \\
				~ & \textsc{GeoMedian} & 0.87 & 0.86 & 0.86 & 0.85 & 0.84 & 0.38 & 0.86 & 0.80 & 0.71 & 0.22 & 0.83 & 0.75 & 0.10 \\
				~ & \textsc{MultiKrum} & 0.88 & 0.12 & 0.86 & 0.80 & 0.10 & 0.12 & 0.86 & 0.73 & 0.10 & 0.16 & 0.85 & 0.69 & 0.10 \\
				~ & \textsc{NNM+Krum} & 0.87 & 0.12 & 0.86 & 0.80 & 0.10 & 0.12 & 0.85 & 0.71 & 0.30 & 0.12 & 0.85 & 0.66 & 0.10 \\ \cmidrule(lr){2-15}
				~ & \textsc{GAN (Fixed Threshold)} & 0.51 & 0.53 & 0.52 & 0.51 & 0.53 & 0.53 & 0.52 & 0.53 & 0.51 & 0.53 & 0.51 & 0.52 & 0.52 \\
				~ & \textsc{GAN (Adaptive: Mean)} & 0.85 & 0.86 & 0.87 & 0.86 & 0.87 & 0.86 & 0.87 & 0.86 & 0.86 & 0.86 & 0.86 & 0.85 & 0.86 \\
				~ & \textsc{GAN (Clustered)} & 0.87 & 0.86 & 0.86 & 0.86 & 0.86 & 0.86 & 0.86 & 0.85 & 0.86 & 0.85 & 0.86 & 0.82 & 0.86 \\
                \midrule[1.125pt]
                \bottomrule[1.5pt]
            \end{tabular}%
        }
    \end{center}
    \footnotesize{*we set $\beta =$ $0.1$, $0.2$, $0.3$ (where $\beta$ is byzantine aggregation heuristic, for \textsc{TrimAvg} and \textsc{MultiKrum} respectively under $10\%$, $20\%$, and $30\%$ malicious clients.}
\end{table*}  % ACC for alpha =  1.0

%%%%%%%%%%%%%%%%%%%%%%%%%%%%%%%%%%%%%%%%%%%%%%%%%%%%%%%%%%%%%%%%%%%%%%%%%%%%%%%%
\end{document}